\newif\ifpreprint

\preprinttrue 

\ifpreprint
\documentclass[aip,jcp,amsmath,amssymb,preprint]{revtex4-1}
\else
\documentclass[aip,jcp,amsmath,amssymb,reprint]{revtex4-1}
\fi

\usepackage{xspace}
\usepackage{graphicx}
\usepackage{braket}
\usepackage[version=4]{mhchem}
\usepackage{multirow}
\usepackage{threeparttable,threeparttablex,booktabs}
\usepackage{longtable}
\usepackage{float}
\usepackage{color}
\usepackage{ifthen}
\usepackage[colorlinks = true,
            linkcolor = blue,
            urlcolor  = black,
            citecolor = blue,
            anchorcolor = black]{hyperref}
\usepackage{algorithm}
\usepackage{algpseudocode}

\newcommand*{\abinitio}{{\it ab initio}\xspace}
\newcommand*{\cm}{cm$^{-1}$\xspace}

\newcommand*{\sunit}{$E_{\rm h}^{-2}$\xspace}
\newcommand*{\Eh}{$E_{\rm h}$\xspace}

\newcommand*{\PSI}{{\scshape Psi4}\xspace}

\newcommand*{\forte}{{\scshape Forte}\xspace}
\newcommand*{\block}{{\scshape Block2}\xspace}

\newcommand{\mref}[0]{{\Psi_0^\alpha}}
\newcommand{\rref}[1]{{\Psi^{\prime}_{#1}}}
\newcommand{\tens}[3]{{#1}_{#2}^{#3}}

\newcommand{\cop}[1]{\hat{a}^\dag_{#1}}
\newcommand{\aop}[1]{\hat{a}_{#1}}

\newcommand{\tdadapt}[2]{\tilde{\Gamma}_{#2}^{#1}}

\newcommand{\cuadapt}[2]{\Lambda_{#2}^{#1}}

\usepackage{newtxtext,newtxmath}
\usepackage[scaled=1.0]{helvet}
\usepackage[compact]{titlesec}
\usepackage{xcolor}
\usepackage{notes2bib}
\usepackage{lineno}

\newcommand*\patchAmsMathEnvironmentForLineno[1]{%
  \expandafter\let\csname old#1\expandafter\endcsname\csname #1\endcsname
  \expandafter\let\csname oldend#1\expandafter\endcsname\csname end#1\endcsname
  \renewenvironment{#1}%
     {\linenomath\csname old#1\endcsname}%
     {\csname oldend#1\endcsname\endlinenomath}}%
\newcommand*\patchBothAmsMathEnvironmentsForLineno[1]{%
  \patchAmsMathEnvironmentForLineno{#1}%
  \patchAmsMathEnvironmentForLineno{#1*}}%
\AtBeginDocument{%
\patchBothAmsMathEnvironmentsForLineno{equation}%
\patchBothAmsMathEnvironmentsForLineno{align}%
\patchBothAmsMathEnvironmentsForLineno{flalign}%
\patchBothAmsMathEnvironmentsForLineno{alignat}%
\patchBothAmsMathEnvironmentsForLineno{gather}%
\patchBothAmsMathEnvironmentsForLineno{multline}%
}

\bibnotesetup{ note-name = , use-sort-key = false}

\titleformat{\section}
  {\normalfont\sffamily\bfseries}
  {\thesection.}{0.5 em}{\MakeTextUppercase}
\titlespacing{\section}{0pt}{12pt}{12pt}

\titleformat{\subsection}[block]
  {\normalfont\sffamily\bfseries}
  {\thesubsection.}{0.5 em}{}
\titlespacing{\subsection}{0pt}{12pt}{8pt}

\titleformat{\subsubsection}[block]
  {\normalfont\itshape\sffamily\bfseries\raggedright}
  {\arabic{subsubsection}.}{0.5 em}{}
\titlespacing{\subsubsection}{0pt}{8pt}{8pt}

\usepackage[normalem]{ulem}

\definecolor{goodorange}{RGB}{225,125,0}
\definecolor{goodgreen}{RGB}{0,125,0}
\definecolor{goodred}{RGB}{220,50,25}
\definecolor{goodblue}{RGB}{25,25,150}

\newcommand{\note}[2]{
\ifthenelse{\equal{#1}{F}}{
\colorbox{goodorange}{\textcolor{white}{\footnotesize \fontfamily{phv}\selectfont #1}}
    \textcolor{goodorange}{{\footnotesize \fontfamily{phv}\selectfont #2}}\xspace
}{}
\ifthenelse{\equal{#1}{Y}}{
\colorbox{goodred}{\textcolor{white}{\footnotesize \fontfamily{phv}\selectfont #1}}
    \textcolor{goodred}{{\footnotesize \fontfamily{phv}\selectfont #2}}\xspace
}{}
}

\usepackage{dcolumn}
\newcolumntype{d}[1]{D{.}{.}{#1}}
\makeatletter
\newcolumntype{B}[3]{>{\boldmath\DC@{#1}{#2}{#3}}c<{\DC@end}}
\makeatother
\makeatletter

\newbox\swb@xone
\newbox\swb@xtwo
\newbox\swb@xthree
\newbox\swb@xfour
\newdimen\swdimen@ne
\newdimen\swdimentw@

\newcommand{\acontraction}[5][1ex]{%
  \mathchoice
    {\acontraction@\displaystyle{#2}{#3}{#4}{#5}{#1}}%
    {\acontraction@\textstyle{#2}{#3}{#4}{#5}{#1}}%
    {\acontraction@\scriptstyle{#2}{#3}{#4}{#5}{#1}}%
    {\acontraction@\scriptscriptstyle{#2}{#3}{#4}{#5}{#1}}}%
\newcommand{\acontraction@}[6]{%
  \setbox\swb@xone=\hbox{${}#1{}#2{}$}%
  \setbox\swb@xtwo=\hbox{${}#1{}#3{}$}%
  \setbox\swb@xthree=\hbox{${}#1{}#4{}$}%
  \setbox\swb@xfour=\hbox{${}#1{}#5{}$}%
  \swdimen@ne=\wd\swb@xtwo%
  \advance\swdimen@ne by \wd\swb@xfour%
  \divide\swdimen@ne by 2%
  \advance\swdimen@ne by \wd\swb@xthree%
  \vbox{%
    \hbox to 0pt{%
      \kern \wd\swb@xone%
      \kern 0.5\wd\swb@xtwo%
      \acontraction@@{\swdimen@ne}{#6}%
      \hss}%
    \vskip 0.5ex
    \vskip\ht\swb@xtwo}}

\newcommand{\acontraction@@}[3][0.05em]{%
  \hbox{%
    \vrule width #1 height 0pt depth #3%
    \vrule width #2 height 0pt depth #1%
    \vrule width #1 height 0pt depth #3%
    \relax}}

\newcommand{\tcontraction}[5][1ex]{%
  \mathchoice
    {\tcontraction@\displaystyle{#2}{#3}{#4}{#5}{#1}}%
    {\tcontraction@\textstyle{#2}{#3}{#4}{#5}{#1}}%
    {\tcontraction@\scriptstyle{#2}{#3}{#4}{#5}{#1}}%
    {\tcontraction@\scriptscriptstyle{#2}{#3}{#4}{#5}{#1}}}%
\newcommand{\tcontraction@}[6]{%
  \setbox\swb@xone=\hbox{${}#1{}#2{}$}%
  \setbox\swb@xtwo=\hbox{${}#1{}#3{}$}%
  \setbox\swb@xthree=\hbox{${}#1{}#4{}$}%
  \setbox\swb@xfour=\hbox{${}#1{}#5{}$}%
  \swdimen@ne=\wd\swb@xtwo%
  \advance\swdimen@ne by \wd\swb@xfour%
  \divide\swdimen@ne by 2%
  \advance\swdimen@ne by \wd\swb@xthree%
  \vbox{%
    \hbox to 0pt{%
      \kern \wd\swb@xone%
      \kern 0.5\wd\swb@xtwo%
      \tcontraction@@{\swdimen@ne}{#6}%
      \hss}%
    \vskip 0.5ex
    \vskip\ht\swb@xtwo}}

\newcommand{\tcontraction@@}[3][0.075em]{%
  \hbox{%
    \vrule width #1 height 0pt depth #3%
    \vrule width #2 height 0pt depth #1%
    \vrule width #1 height 0pt depth #3%
    \relax}}

\makeatother

\begin{document}

\title{Driven similarity renormalization group with a large active space: Applications to oligoacenes, zeaxanthin, and chromium dimer}

\author{Chenyang Li}
\email{chenyang.li@bnu.edu.cn}
\affiliation{Key Laboratory of Theoretical and Computational Photochemistry, Ministry of Education, College of Chemistry, Beijing Normal University, Beijing 100875, China}

\author{Xiaoxue Wang}
\affiliation{Key Laboratory of Theoretical and Computational Photochemistry, Ministry of Education, College of Chemistry, Beijing Normal University, Beijing 100875, China}

\author{Huanchen Zhai}
\email{hzhai@flatironinstitute.org}
\affiliation{Initiative for Computational Catalysis, Flatiron Institute, 160 5th Avenue, New York, NY 10010, USA}

\author{Wei-Hai Fang}
\affiliation{Key Laboratory of Theoretical and Computational Photochemistry, Ministry of Education, College of Chemistry, Beijing Normal University, Beijing 100875, China}

\date{\today}

\begin{abstract}
We present a new implementation of the driven similarity renormalization group (DSRG) based on a density matrix renormalization group (DMRG) reference.
The explicit build of high-order reduced density matrices is avoided by forming matrix-product-state compressed intermediates.
This algorithm facilitates the application of DSRG second- and third-order perturbation theories to dodecacene with an active space of 50 electrons in 50 orbitals.
This active space appears the largest employed to date within the framework of internally contracted multireference formalism.
The DMRG-DSRG approach is applied to several challenging systems, including the singlet-triplet gaps ($\Delta_{\rm ST}$) of oligoacenes ranging from naphthalene to dodecacene, the vertical excitation energies of zeaxanthin, and the ground-state potential energy curve (PEC) of \ce{Cr2} molecule.
Our best estimate for the vertical $\Delta_{\rm ST}$ of dodecacene is 0.22~eV, showing an excellent agreement with that of the linearized adiabatic connection method (0.24~eV).
For zeaxanthin, all DSRG schemes suggest the order of $\rm 2\, ^1 A_g^- < 1\, ^1 B_u^+ < 1\, ^1 B_u^-$ for excited states.
Both the equilibrium and the shoulder regions of the \ce{Cr2} PEC are reasonably reproduced by the linearized DSRG with one- and two-body operators.
\end{abstract}

\maketitle

\section{Introduction}
\label{sec:intro}

An accurate description of systems involving strongly correlated electrons continues to be one of the great challenges in \abinitio quantum chemistry.
The wave function of such systems exhibits notable deviations from the mean-field picture and typically necessitates a linear combination of electron configurations for qualitatively correct results.
While it is now possible to exactly solve the ground state for 24 electrons in 24 molecular orbitals (MOs) using a wave function composed of $10^{12}$ Slater determinants,\cite{Vogiatzis:2017gp} the cost is prohibitive when applying a similar treatment to more MOs.
Numerous approximate approaches have been introduced to go beyond this limit, such as density matrix renormalization group (DMRG),\cite{White:1992ie,Schollwock:2011gl,Chan:2016fg} selected configuration interaction (sCI),\cite{Evangelisti1983,Schriber:2016kl,Holmes:2016fm,Liu:2016jwa} full configuration interaction quantum Monte Carlo,\cite{Booth:2009hb} and the method of variational two-electron reduced density matrix (v2RDM).\cite{Gidofalvi2008,FossoTande:2016hb}
Some of these approaches have enabled near-exact computations for systems with more than 100 orbitals.\cite{Sharma:2017iu,Zhai2021}
Nevertheless, it remains insufficient for broader applications in quantum chemistry.

Multireference (MR) methods offer a more practical route for molecular systems by making a distinction between static and dynamical electron correlation.
The static correlation arises from the strong mixing of quasidegenerate configurations in the active space and requires a near-exact treatment using the methods mentioned above.
These configurations of the active space are generated by allocating a number of electrons to a set of active orbitals while keeping the core and virtual orbitals doubly occupied and unoccupied, respectively.
The resulting wave function provides a decent reference state for the correction of dynamical correlation, which is captured by particle-hole excitations using either configuration interaction (MRCI),\cite{Szalay:2012df,Luo:2018gn} perturbation theory (MRPT),\cite{Andersson:1992cq,Angeli:2001bg,Sharma:2017eo} or coupled cluster theory (MRCC).\cite{Lyakh:2012cn,Evangelista:2018bt}
We note that both static and dynamical correlation should be addressed in order to achieve high accuracy, especially for the excited states of conjugated systems\cite{Sokolov:2017gr} and the potential energy curve of chromium dimer.\cite{Andersson1994,Larsson2022}

The second-order MRPT represents a collection of affordable and extensively employed MR approaches, including the complete active space perturbation theory (CASPT2)\cite{Andersson:1992cq} and \textit{n}-electron valence state perturbation theory (NEVPT2).\cite{Angeli:2001bg}
Both CASPT2 and NEVPT2 techniques have been integrated with DMRG to effectively address large active spaces.\cite{Kurashige:2011ck,Guo:2016fu}
However, the necessity for high-order reduced density matrices (RDMs) constrains their applicability to approximately 30 active orbitals.\cite{Kurashige:2014bq,Guo:2016fu,Phung:2016ke}
Alternative MR methods that rely solely on 1- and 2-RDMs have been proposed.
Notable examples are multiconfiguration pair-density functional theory (PDFT),\cite{LiManni:2014kg} and the linearized adiabatic connection method (AC0),\cite{Pernal:2018fg,Pastorczak:2018fl} both of which have shown applicable for more than 45 active orbitals using either a generalized active space (GAS)\cite{Ghosh:2017gl} or a DMRG reference wave function.\cite{Zuzak2024}

The driven similarity renormalization group (DSRG) provides another promising route to computations for large active spaces.\cite{Evangelista:2014kt,Li:2019fu}
The MR-DSRG approach is closely related to in-medium similarity renormalization group,\cite{Hergert:2014je} canonical transformation theory,\cite{Yanai:2006gi,Yanai:2007ix} and internally contracted MRCC methods.\cite{Datta:2011ca,Hanauer:2011ey}
In MR-DSRG,\cite{Li:2019fu} the dynamical correlation is addressed by a unitary transformation to the Hamiltonian and a parametrized flow equation is adopted to mitigate numerical instabilities that may arise in other internally contracted formulations.\cite{Yanai:2007ix,Datta:2011ca,Hanauer:2011ey}
With the linearized commutator approximation,\cite{Yanai:2006gi,Li:2016hb} the resulting MR-LDSRG(2) energy expression incorporates at most 3-RDMs.
A perturbative analysis on the MR-LDSRG(2) equations leads to the development of the DSRG second- and third-order perturbation theories (PT2/3).\cite{Li:2015iz,Li:2017bx}
Additionally, a state-averaged (SA) extension of DSRG has been introduced for electronic excited states,\cite{Li:2018kl} where only 1- and 2-RDMs are necessary for vertical transition energies.
The (SA-)DSRG hierarchy has been benchmarked on second-row diatomic molecules,\cite{Zhang:2019wj,Li:2021sf} transition-metal energetics,\cite{Li:2021sf,Phung2023} and excited states,\cite{Wang2023,Huang2022d,Huang2024b} showing comparable accuracy to other well-established methods.

In this work, we extend the (SA-)DSRG methodology by adopting DMRG as the active space solver to facilitate applications that necessitate large active spaces.
The (SA-)DSRG-PT2 based on a DMRG or sCI reference wave function has been reported and applied to oligoacenes,\cite{Schriber:2018hw} polyenes,\cite{Khokhlov2021} and carotenoids.\cite{Khokhlov2022}
However, the number of active orbitals is limited to 30 in these studies, which is well below the potential of DSRG-PT2 for incorporating large active spaces.\cite{Li:2017ff}
Herein, we implement an algorithm based on the compressed representation of intermediate states\cite{Sokolov:2017gr} to avoid the explicit computation of 3-RDMs of the reference, broadening the scope of DMRG-DSRG-PT2/3 to accommodate up to 50 active orbitals.
Furthermore, the non-perturbative DMRG-LDSRG(2) method is applied to nonacene with an active space of 38 electrons in 38 orbitals, marking the largest active space employed in a MRCC computation to date.

\begin{figure}[h]
\centering
\includegraphics[width=0.6\columnwidth]{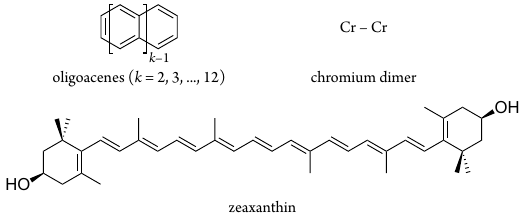}
\caption{Molecular systems studied in this work.}
\label{fig:mols}
\end{figure}

We apply the DMRG-DSRG methods to oligoacenes, zeaxanthin, and chromium dimer, as shown in Fig.~\ref{fig:mols}.
These three molecules exemplify some of the challenging strongly correlated systems for \abinitio quantum chemistry concerning polyradicals, excited states of conjugated molecules, and transition-metal chemistry.
Oligoacenes consist of linearly fused benzene rings and only one Clar sextet can be drawn among all resonance structures, suggesting an increase in reactivity as the size grows.\cite{Tonshoff2021}
Indeed, both theory and experiment support this argument.
Various theoretical methods suggest the emergence of polyradical character in large oligoacenes,\cite{Hachmann:2007ft,Yang:2016by,FossoTande:2016hb,Trinquier2018} characterized by the small singlet-triplet splittings ($\Delta_{\rm ST}$).
The synthesis of oligoacenes longer than hexacene is only possible under matrix isolation conditions or on-surface generations.\cite{Mondai2006,Shen2018,Zuzak2018,Eisenhut2020,Zuzak2024,Ruan2025}
The largest oligoacene observed experimentally to date is pentadecacene ($k = 15$), with a $\Delta_{\rm ST}$ of 124~meV.\cite{Ruan2025}
Nonetheless, accurately predicting the $\Delta_{\rm ST}$ of large oligoacenes appears challenging for theories.\cite{Zuzak2024}

Zeaxanthin is a natural carotenoid commonly found in the xanthophyll cycle of plants and retinas of animal eyes.
The extended chain of alternating single and double bonds leads to a unique electronic structure for the excited state, which closely resembles that of polyenes in $\rm C_{2h}$ symmetry.\cite{Tavan:1987gc}
It is generally accepted that a dark state $\rm 2\, ^1A_g^-$ exists between the optically accessible $\rm 1\, ^1B_u^+$ state and the ground state ($\rm 1\, ^1A_g^-$).\cite{Kopczynski2007,Ostroumov2013}
The presence of an additional weakly allowed transition lying below the $\rm 1\, ^1B_u^+$ state has also been suggested by experiments for carotenoids with nine or more double bonds.\cite{Wang2005,Ostroumov2013}
However, the relative vertical energies of these states remain controversy for theories,\cite{Andreussi2015,Khokhlov:2020kn,Khokhlov2022} as they cannot be easily extracted from experiments.
Here, we provide SA-DSRG predictions on the vertical excitation energies of zeaxanthin, which in principle surpass the existing applications of density function theory (DFT) and second-order MRPTs.\cite{Gotze2013,Andreussi2015,Khokhlov2022}

Chromium dimer is one of the most notorious diatomic molecules for molecular electronic-structure theories.\cite{Scuseria1990,Andersson1994,Larsson2022}
The unusual bonding character along the dissociation coordinates poses a significant issue for theoretical computations and requires a balanced treatment between static and dynamical electron correlation.\cite{Roos:1995jz,Angeli:2006gf,Muller:2009jd,Kurashige:2011ck,Guo:2016fu,Li:2019wp,Larsson2022}
Extensive theoretical attempts have been made to reproduce the experimental curve,\cite{Casey:1993gq} which is achieved very recently using a composite approach of DMRG, sCI, and matrix-product-state-approximated multireference retaining excitation PT2 (MPS-REPT2).\cite{Sharma:2017eo,Larsson2022}
Despite this tremendous feat, the MR-DSRG prediction on this challenging system remains unexplored and it may serve as a serious benchmark for MR-DSRG on transition-metal complexes.

This article is organized as follows.
Section~\ref{sec:theory} provides a concise overview of the DSRG and DMRG theories, followed by a discussion on their integration and the associated challenges when using large active spaces.
The DMRG-DSRG approach is employed to compute the singlet-triplet gaps of oligoacenes, the vertical transition energies of zeaxanthin, and the ground-state potential energy curve of chromium dimer.
The computational details and numerical results are presented in Secs.~\ref{sec:compt} and \ref{sec:results}, respectively.
Lastly, we conclude and finish our discussions in Sec.~\ref{sec:conclusion}.

\section{Theory}
\label{sec:theory}

\subsection{State-averaged driven similarity renormalization group (SA-DSRG)}
\label{sec:dsrg}

We begin with an overview of the SA-DSRG approach and more details can be found in Refs.~\citenum{Li:2018kl} and \citenum{Li:2019fu}.
The SA-DSRG ansatz incorporates the dynamical electron correlation on top of a set of zeroth-order multi-configuration reference states $\{ \mref{\alpha}| \alpha = 1, 2, \cdots, n \}$ by forming an effective Hamiltonian $\bar{H}(s)$ via a unitary transformation to the Born-Oppenheimer Hamiltonian $\hat{H}$
\begin{align}
\label{eq:dsrg}
\hat{H} \rightarrow \bar{H}(s) = \hat{U}^\dag(s) \hat{H} \hat{U}(s).
\end{align}
Here, the parameter $s \in [0,\infty)$ controls the extent of the transformation such that excitations with energy denominators smaller than $s^{-1/2}$ are dampened, effectively mitigating the intruder-state problem.\cite{Roos:1995jz}
Specifically, we require the many-body components of $\bar{H}(s)$ [$\tens{\bar{H}}{ab\cdots}{ij\cdots} (s)$] equal to the first-order Hamiltonian elements of the single-reference similarity renormalization group [$\tens{r}{ab\cdots}{ij\cdots} (s)$]
\begin{align}
\label{eq:flow}
\tens{\bar{H}}{ab\cdots}{ij\cdots} (s) = \tens{r}{ab\cdots}{ij\cdots} (s),
\end{align}
where indices $i,j,\dots$ label hole (core and active) and $a,b,\dots$ label particle (active and virtual) orbitals.

In the linear two-body approximation of DSRG [LDSRG(2)],\cite{Li:2016hb} the unitary transformation $\hat{U}(s)$ is parametrized using the one- [$\hat{T}_1(s)$] and two-body [$\hat{T}_2(s)$] cluster operators
\begin{align}
\hat{U}(s) = e^{\hat{A}_1(s) + \hat{A}_2 (s)}, \quad \hat{A}_k(s) = \hat{T}_k(s) - \hat{T}_k^\dagger(s),
\end{align}
and the transformed Hamiltonian $\bar{H}(s)$ is approximated by keeping at most two-body terms for every commutator in the Baker--Campbell--Hausdorff (BCH) expansion
\begin{align}
\label{eq:Hbar_L12}
\bar{H}(s) \approx&\, \hat{H} + [\hat{H}, \hat{A}_{1,2}(s)]_{0,1,2} \notag\\
&+ \frac{1}{2} [ [\hat{H}, \hat{A}_{1,2}(s)]_{1,2}, \hat{A}_{1,2}(s) ]_{0,1,2} + \cdots
\end{align}
with the short-hand notation $\hat{A}_{1,2}(s) = \hat{A}_1(s) + \hat{A}_2 (s)$.
Equation \eqref{eq:flow} can now be understood as the set of equations to determine the cluster operators, solving which requires an iterative procedure in a similar fashion to the coupled cluster theory with singles and doubles.
By contrast, the perturbation theory formulated by truncating the LDSRG(2) $\bar{H}(s)$ to second- and third-order terms (DSRG-PT2/3) is non-iterative.\cite{Li:2015iz,Li:2017bx}

Once the cluster operators are determined, we solve the active-space eigenvalue problem using $\bar{H}(s)$ for the ground- and excited-state energies [$E_{\alpha} (s)$] and the relaxed reference states [$\rref{\alpha} (s)$]:
\begin{equation}
\label{eq:dsrg_relax}
\bar{H}(s) \ket{\rref{\alpha} (s)} = E_{\alpha} (s) \ket{\rref{\alpha} (s)}.
\end{equation}
In this work, both the original ($\mref{\alpha}$) and the relaxed [$\rref{\alpha} (s)$] reference states are computed using DMRG (see Sec.~\ref{sec:mps}).
The relaxed states $\rref{\alpha} (s)$ may also be taken as a new set of reference states for DSRG, leading to alternating solutions between Eq.~\eqref{eq:dsrg_relax} and Eq.~\eqref{eq:flow}.
The converged energies resulting from this iterative procedure are referred to as the fully relaxed energies.\cite{Li:2017bx}
All LDSRG(2) results in this work correspond to those of fully relaxed.
By contrast, a one-shot relaxation step [Eq.~\eqref{eq:dsrg_relax}] is performed for DSRG-PT2/3 due to a balance between cost and accuracy.\cite{Li:2017bx}

We note that evaluating the commutators of Eq.~\eqref{eq:Hbar_L12} requires the 1-, 2-, and 3-particle RDMs of the reference states.
Importantly, the 3-RDM contributions only appear in the scalar terms (e.g., $[\hat{H}, \hat{A}_{1,2}(s)]_{0}$).
As such, there is no need to compute the 3-RDMs for vertical transition energies in the SA-DSRG formalism.

\subsection{Density matrix renormalization group (DMRG)}
\label{sec:mps}

As mentioned previously, we adopt the DMRG approach to account for the static correlation within the active space and to obtain the RDMs required for evaluating the DSRG Hamiltonian.
To this end, we write a reference state $\mref{}$ in the form of matrix product state (MPS):
\begin{align}
\label{eq:mps}
\ket{\mref{}} = \sum_{n_1 n_2 \cdots n_L} \mathbf{A}[1]^{n_1} \mathbf{A}[2]^{n_2} \cdots \mathbf{A}[L]^{n_L} \ket{n_1 n_2 \dots n_L},
\end{align}
where $L$ is the number of active orbitals (sites) and $n_l$ labels the occupation state of site $l$.
Every matrix ${\bf A}[l]$ with $1 < l < L$ is of dimension $M \times M$, while the leftmost and rightmost matrices are $1 \times M$ and $M \times 1$ vectors, respectively.
The parameter $M$ is the so-called MPS bond dimension, which controls the accuracy and complexity of the DMRG algorithm.
In Eq.~\eqref{eq:mps} and the following, we drop the state label $\alpha$ in the reference state for brevity.

The active space eigenvalue problem [e.g., Eq.~\eqref{eq:dsrg_relax}] is equivalent to minimizing the Rayleigh quotient ${\cal R} = \braket{ \mref{} | \hat{H} | \mref{} } / \braket{ \mref{} | \mref{} }$ for the Hamiltonian $\hat{H}$.
Here, $\hat{H}$ is also in the matrix-product form, termed matrix product operators (MPOs).\cite{Schollwock:2011gl,Chan:2016fg}
The minimization of $\cal R$ is achieved via the sweep algorithm,\cite{White:1993fb} which optimizes the MPS matrices step by step.
We adopt the state-averaged representation for solving the ground and excited states.\cite{Dorando2007}
Briefly, we find $n$ orthogonal states in a common basis of renormalized states constructed at a given site during the sweep.
The density matrices of these states are then averaged to build the common renormalized basis in the next iteration.
The computational and memory complexities of the state-averaged DMRG are $O(nL^3M^3 + L^4M^2)$ and $O(L^2M^2)$, respectively.
We assume the spin-adapted formalism\cite{Sharma2012,Wouters:2014gl,Keller2016} in all DMRG computations.
Further details on \abinitio DMRG can be found in Refs.~\citenum{Chan:2016fg}, \citenum{Zhai2021} and \citenum{Zhai2023}.

Given an optimized MPS \(\ket{\mref{}}\) with bond dimension \(M\), the spin-traced RDMs are computed using a DMRG-like sweep algorithm.
At every iteration during a sweep, we obtain a few RDM elements via tensor contractions and reuse the partially contracted intermediates for efficiency.\cite{Kurashige:2014bq,Guo:2016fu}
For each state, the complexity of building $k$-RDM scales as $O(L^{k+1}M^3+L^{2k}M^2)$ in computational time\cite{Guo:2016fu} and $O(L^kM^2)$ in memory.
We note that algorithms with lower memory requirements exist, yet a larger prefactor emerges in the computational cost and the cost of writing scratch files.\cite{Zhai2023}

\subsection{SA-DSRG based on DMRG reference wave functions}
\label{sec:comb}

The combination of SA-DSRG and DMRG is straightforward.
We employ DMRG to obtain wave functions within the active space and subsequently correct for the missing dynamical correlation using DSRG.
The key ingredients can be summarized as follows.
Given an ordering of the active orbitals, the zeroth-order reference states of SA-DSRG are obtained by the SA-DMRG self-consistent field (SCF), where both MPSs and molecular orbitals are simultaneously optimized.
Then, the 1-, 2-, and 3-RDMs of the converged MPSs are computed.
However, the 3-RDM can be safely ignored without affecting the vertical transition energies as stated in Sec.~\ref{sec:dsrg}.
At this stage, we can build the DSRG Hamiltonian and solve for the cluster amplitudes for a given value of $s$ [Eq.~\eqref{eq:flow}].
Finally, the active-space eigenvalue problem with the DSRG transformed Hamiltonian is solved using DMRG to obtain the relaxed energies [Eq.~\eqref{eq:dsrg_relax}].

When targeting large active spaces (e.g., $L > 45$), the direct integration between DMRG and DSRG may run into memory issue due to the computation of 3-RDM.
By itself, there are $L^6$ elements in 3-RDM and it may still fit in memory for $L \sim 50$.
However, a computationally efficient evaluation of 3-RDMs in DMRG requires a memory allocation of $O(L^3 M^2)$.\cite{Kurashige:2014bq}
This memory demand quickly becomes the bottleneck for any decent DMRG computations with $M \sim 1000$.
To this end, several strategies have been proposed to alleviate the issue of high-order RDMs in internally contracted MR theories.\cite{Zgid:2009fu,Kurashige:2014bq,Li2023a,Chatterjee2020,Kollmar2021}

These approaches roughly fall into two categories.
One is to approximate the high-order RDMs using the cumulant decomposition and then entirely or partially ignore the high-order density cumulants.\cite{Zgid:2009fu,Kurashige:2014bq,Li2023a}
For example, Evangelista and co-workers\cite{Li2023a} suggest a partial neglect scheme of keeping the elements of the 3-body density cumulant ($\boldsymbol \Lambda_3$) with at least one identical particle-hole pair in the semicanonical basis:
\begin{align}
\label{eq:L3d}
\boldsymbol \Lambda_3^{\rm d} =
\begin{cases}
\cuadapt{xyw}{uvw}, \cuadapt{xwy}{uvw}, \cuadapt{wxy}{uvw}, \dots\\
0, \quad \text{otherwise}.
\end{cases}
\end{align}
The error introduced to the DSRG-PT2 energy due to Eq.~\eqref{eq:L3d} is found less than 1~m\Eh.\cite{Li2023a}
Due to the index permutation symmetry of $\boldsymbol \Lambda_3$, only $\cuadapt{xyw}{uvw}$ and $\cuadapt{xwy}{uvw}$ need to be stored.
These two quantities must be computed directly from the diagonal 3-RDMs and low-order RDMs.
However, Eq.~\eqref{eq:L3d} is not invariant with respect to orbital rotations.
This aspect is important in DMRG-DSRG because DMRG performs the best using localized active orbitals (particularly for the quasi-one-dimensional systems studied in this work), while DSRG assumes semicanonical orbitals.
To obtain the semicanonical $\boldsymbol \Lambda_3^{\rm d}$, one needs to perform orbital rotations to \emph{all} six indices of $\boldsymbol \Lambda_3$ in localized basis, rendering this approach not applicable for our purposes.

The other category is to avoid the explicit build of high-order RDMs by forming clever intermediates for the impeding terms.\cite{Sokolov:2017gr,Chatterjee2020,Kollmar2021}
In this work, we adopt the algorithm of MPS compressed intermediates as introduced in time-dependent NEVPT2.\cite{Sokolov:2017gr}
Take one of the 3-RDM contributions to the DSRG energy as an example:
\begin{align}
[\hat{H}, \hat{A}_{2}(s)]_{0} &\leftarrow \sum_{e}^{\rm virtual} \sum_{uvwxyz}^{\rm active} \braket{ew | xy} \tens{t}{ez}{uv} (s) \tdadapt{xyz}{uwv}, \label{eq:H2T2C0v} \\
\tdadapt{xyz}{uwv} &= \sum_{\sigma\sigma'\sigma''}^{\rm spin} \braket{\Psi | \cop{x\sigma} \cop{y\sigma'} \aop{w\sigma'} \cop{z\sigma''} \aop{v\sigma''} \aop{u\sigma} | \Psi }, \label{eq:SFG3}
\end{align}
where $\braket{ew|xy}$, $\tens{t}{ez}{uv} (s)$, and $\cop{p}$ ($\aop{p}$) are two-electron integrals, cluster amplitudes, and creation (annihilation) operators, respectively.
Instead of storing $\tdadapt{xyz}{uwv}$, one may evaluate Eq.~\eqref{eq:H2T2C0v} by first building an intermediate state for every virtual orbital
\begin{align}
\label{eq:ecps}
\ket{\Phi^{e}_{\sigma}} &= \sum_{uvz}^{\rm active} \sum_{\sigma'}^{\rm spin} \cop{z\sigma'} \aop{v\sigma'} \aop{u\sigma} \ket{\Psi } \tens{t}{ez}{uv} (s)
\end{align}
and rewrite Eq.~\eqref{eq:H2T2C0v} as
\begin{align}
\label{eq:ecps-expt}
[\hat{H}, \hat{A}_{2}(s)]_{0} &\leftarrow \sum_{e}^{\rm virtual} \sum_{wxy}^{\rm active} \sum_{\sigma\sigma'}^{\rm spin} \braket{\Phi^{e}_{\sigma}| \cop{w\sigma'} \aop{y\sigma'} \aop{x\sigma} | \Psi } \braket{ew | xy}.
\end{align}
The intermediate state $\ket{\Phi^{e}_{\sigma}}$ can be exactly expressed as an MPS of bond dimension $LM$.
We truncate this bond dimension to $M'$ via a variational minimization of the distance between the approximate state $\ket{\tilde{\Phi}^{e}_{\sigma}}$ and $\ket{\Phi^{e}_{\sigma}}$ using a DMRG-like sweep algorithm.
For every virtual index, the bond dimension of the required MPO [e.g., $\sum_{uvz}^{\rm active} \sum_{\sigma'}^{\rm spin} \tens{t}{ez}{uv} (s) \cop{z\sigma'} \aop{v\sigma'} \aop{u\sigma}$] is $L$ for both Eqs.~\eqref{eq:ecps} and ~\eqref{eq:ecps-expt}, leading to a $O(L^2 M'^2 M + L^3M'M)$ cost in computation with a $O(L M'M)$ memory requirement.
Therefore, the memory cost is reduced by at least a factor of $L$ compared to the standard 3-RDM algorithm.
The computational complexity may also benefit from this algorithm if $M' < \sqrt{\frac{M + L^2}{VM}} LM$ with $V$ being the number of virtual orbitals.
We will present the accuracy of this approximation numerically in Sec.~\ref{sec:ployacene}.
We also mention that this approach can be efficiently combined with the recursive evaluation of the BCH expansion [Eq.~\eqref{eq:Hbar_L12}] because $\ket{\tilde{\Phi}^{e}_{\sigma}}$ can be reused for all nested commutators.

\section{Computational Details}
\label{sec:compt}

We implemented an interface between open-source codes \block\cite{Zhai2023} and \forte.\cite{Evangelista2024}
The former provides efficient algorithms for DMRG and RDMs computations, while various DSRG methods are available in the latter.
Special treatments of the 3-RDMs (Sec.~\ref{sec:comb}) were implemented in a local branch of \forte.\cite{MyForte}

For oligoacenes, we directly took the geometries from Ref.~\citenum{Ghosh:2017gl}, which were optimized using B3LYP/6-31G(d,p).
We considered the complete valence $\pi$ orbitals as active, resulting in ($4k+2$,$4k+2$) active space for $k$-acene.
As usual, the ($N$,$L$) pair indicates an active space of $N$ electrons in $L$ orbitals.
Both adiabatic and vertical singlet-triplet gaps ($\Delta_{\rm ST} = E_{\rm T_1} - E_{\rm S_0}$) were reported.
The adiabatic $\Delta_{\rm ST}$ was calculated as the energy difference of two state-specific (SS) DMRG-DSRG computations using individually optimized orbitals and geometries.
Contrarily, the vertical $\Delta_{\rm ST}$ was obtained by one single SA-DMRG-DSRG computation using the ground-state geometry and the SA-DMRG-SCF orbitals averaged over both states with equal weights.

The above SA-DMRG-DSRG procedure was also employed to compute the vertical excitation energies of zeaxanthin.
Following Ref.~\citenum{Khokhlov2022}, we adopted the (22,22) active space for the entire $\pi$ system.
The lowest three singlet states ($\rm 1\, ^1A_g^-, 2\, ^1A_g^-, 1\, ^1B_u^-$) were averaged in SA-DMRG-SCF, while the $\rm 1\, ^1B_u^+$ state was additionally included in the subsequent SA-DSRG.
The ground-state ($\rm 1\, ^1A_g^-$) geometry of zeaxanthin was taken from Ref.~\citenum{Khokhlov2022}, which was optimized at the CAM-B3LYP/cc-pVDZ level of theory.

The pseudo-one-dimension structure of oligoacenes and zeaxanthin enabled us to use a small bond dimension in DMRG computations.
In consistent with previous findings,\cite{Sharma2019,Khokhlov2022} we found that $M = 600$ was sufficient to yield converged energy differences and therefore, this value was used in all DMRG computations of oligoacenes and zeaxanthin.
The initial guess of active orbitals for DMRG-SCF was obtained by the Pipek-Mezey localization\cite{Pipek:1989ci} sorted in the Fiedler ordering.\cite{OlivaresAmaya:2015hi}

For chromium dimer, we tested two active spaces: (12,12) and (12,22).
The former was derived from the 3d and 4s orbitals of Cr atoms, while the latter was obtained by augmenting the (12,12) active space with an additional set of d orbitals.
Following an earlier work,\cite{Kurashige:2009gs} we ordered the 22 active orbitals by the bonding characters: $\sigma, \sigma^\ast, \pi_{xz}, \pi^\ast_{xz}, \pi_{yz}, \pi^\ast_{yz}, \delta_{xy}, \delta^\ast_{xy}, \delta_{x^2-y^2}, \delta^\ast_{x^2-y^2}$.
The DMRG bond dimension was set to $M = 1000$ for a balance between cost and accuracy (discarded weights $\leq 5\times10^{-6}$).
We note that the conventional exactly diagonalized complete active space (CAS) was adopted for the smaller (12,12) active space.
The scalar relativistic effect was described using the second-order Douglas--Kroll--Hess Hamiltonian.\cite{Hess:1986cz,Wolf:2002ia}
The final single-point energy was extrapolated to the complete basis set (CBS) limit $E^{\infty} = E_{\rm ref}^{\infty} + E_{\rm corr}^{\infty}$, which consists of the DMRG reference energy ($E_{\rm ref}^{\infty}$) and the DSRG correlation energy ($E_{\rm corr}^{\infty}$):\cite{Feller:1993ex,Helgaker:1997bb}
\begin{align}
E_{\rm ref} (X) &= E_{\rm ref}^{\infty} + a \exp(-b X), \label{eq:cbs3point} \\
E_{\rm corr} (X) &= E_{\rm corr}^{\infty} + c X^{-3}. \label{eq:cbs2point}
\end{align}
Here, $E (X)$ is the energy of the cc-pwCV$\bar{X}$Z-DK basis set\cite{Balabanov:2005hm} with $X = 3, 4, 5$ for $\bar{X} = {\rm T, Q, 5}$, respectively.
Specifically, we adopted $X = 4$ and $5$ to determine the fitting parameters in Eq.~\eqref{eq:cbs2point}.

All computations utilized the density-fitted (DF) two-electron integrals from \PSI,\cite{Smith:2020ci} where a common auxiliary basis set was used for the entire DMRG-DSRG procedure.
We employed the def2-universal-JKFIT auxiliary basis set\cite{Pritchard:2019gs} for oligoacenes and zeaxanthin, while the auxiliary basis set for \ce{Cr2} was generated by the AutoAux method.\cite{Stoychev2017,Pritchard:2019gs}
The frozen-core approximation was assumed for all DSRG treatments of electron correlation.

In this work, we only considered the sequential version of LDSRG(2) [sq-LDSRG(2)],\cite{Zhang:2019wj} which has been shown to yield similar accuracy to that of LDSRG(2).\cite{Zhang:2019wj,Li:2021sf}
In sq-LDSRG(2), the Hamiltonian is computed as $\bar{H}(s) = e^{-\hat{A}_2 (s)} [e^{-\hat{A}_1 (s)} \hat{H} e^{\hat{A}_1 (s)}] e^{\hat{A}_2 (s)}$, where the intermediate in the bracket $\tilde{H}(s) = e^{-\hat{A}_1 (s)} \hat{H} e^{\hat{A}_1 (s)}$ can be firstly evaluated via orbital rotations and the linear two-body approximation [Eq.~\eqref{eq:Hbar_L12}] only applies to $e^{-\hat{A}_2 (s)} \tilde{H}(s) e^{\hat{A}_2 (s)}$.
Moreover, for each commutator in Eq.~\eqref{eq:Hbar_L12} the two-body terms labeled by three and four virtual indices were ignored.\cite{Zhang:2019wj}
We truncated the virtual orbital space by keeping 98\% of the total DSRG-PT2 natural occupation numbers\cite{Li2024c} to reduce the computational cost for the DSRG-PT3 and sq-LDSRG(2) computations of oligoacenes and zeaxanthin.
Unless otherwise noted, the flow parameter was set to 0.5~\sunit, which were shown to yield accurate potential energy surfaces.\cite{Li:2015iz,Li:2017bx,Li:2018kl}
For vertical transition energies, we used the previously calibrated values for SA-DSRG-PT3 ($s = 2.0$~\sunit) and SA-sq-LDSRG(2) ($s = 1.5$~\sunit).\cite{Wang2023}

\section{Numerical Results}
\label{sec:results}

\subsection{Singlet-triplet splittings of oligoacenes}
\label{sec:ployacene}

\begin{table*}[h]
\begin{threeparttable}

\scriptsize
\renewcommand{\arraystretch}{1.15}
\caption{Vertical singlet-triplet splittings (in eV) of the $k$-acene ($k = 2, 3, \dots, 12$) series.}
\label{tab:acene_vert}

\begin{tabular*}{\textwidth}{@{\extracolsep{\stretch{1}}} c c *{11}{d{1.2}} @{}}
\hline
\hline
& & \multicolumn{5}{c}{SA-DMRG-DSRG\tnote{a}} \\
\cline{3-7}
& & \multicolumn{3}{c}{def2-SVP} & \multicolumn{2}{c}{def2-TZVP} & \multicolumn{1}{c}{GAS-} & \multicolumn{3}{c}{DMRG-} \\
\cline{3-5} \cline{6-7} \cline{9-11}
$k$ & active space & \multicolumn{1}{c}{PT2} & \multicolumn{1}{c}{PT3} & \multicolumn{1}{c}{sq-L2} & \multicolumn{1}{c}{PT2} & \multicolumn{1}{c}{PT3}  & \multicolumn{1}{c}{PDFT\tnote{b}} & \multicolumn{1}{c}{PDFT\tnote{c}} & \multicolumn{1}{c}{AC0\tnote{d}} & \multicolumn{1}{c}{ec-MRCISD+Q\tnote{e}} & \multicolumn{1}{c}{pp-RPA\tnote{f}} & \multicolumn{1}{c}{CCSD(T)\tnote{g}} \\
\hline

2 & (10,10) & 3.13 & 3.17 & 3.19 & 3.10 & 3.15 & 3.25 & 3.35 & 3.36 & 3.43 & 2.87 & 3.30 \\
3 & (14,14) & 2.22 & 2.31 & 2.31 & 2.18 & 2.27 & 2.19 & 2.33 & 2.44 & 2.48 & 1.98 & 2.46 \\
4 & (18,18) & 1.58 & 1.69 & 1.68 & 1.55 & 1.65 & 1.50 & 1.58 & 1.78 & 1.82 & 1.39 & 1.75 \\
5 & (22,22) & 1.15 & 1.26 & 1.25 & 1.11 & 1.22 & 1.08 & 1.13 & 1.33 & 1.36 & 0.98 & 1.36 \\
6 & (26,26) & 0.80 & 0.91 & 0.90 & 0.76 & 0.86 & 0.75 & 0.79 & 1.01 & 0.98 & 0.66 & 0.99 \\
7 & (30,30) & 0.52 & 0.60 & 0.60 & 0.48 & 0.56 & 0.46 & 0.61 & 0.78 & 0.68 & 0.39 & 0.78 \\
8 & (34,34) & 0.37 & 0.43 & 0.43 & 0.33 & 0.39 & 0.23 &  & 0.42 & 0.49 & 0.23 & 0.58 \\
9 & (38,38) & 0.29 & 0.34 & 0.33 & 0.25 & 0.29 & 0.19 &  & 0.33 & 0.37 & 0.15 & 0.46 \\
10 & (42,42) & 0.24 & 0.28 &  & 0.21 & 0.24 & 0.16 &  & 0.27 &  & 0.11 & 0.35 \\
11 & (46,46) & 0.22 & 0.26 &  & 0.19 & [0.23] & 0.07 &  & 0.24 &  & 0.11 & 0.31 \\
12 & (50,50) & 0.20 & 0.24 &  & 0.18 & [0.22] &  &  & 0.24 &  & 0.13 &  \\

\hline
\hline
\end{tabular*}

\begin{tablenotes}
\item [a] This work, UB3LYP/6-31G(d,p) geometries from Ref.~\citenum{Ghosh:2017gl}. The DSRG-PT3 values in brackets are estimated using the additive assumption $\Delta_{\rm ST}^\text{PT3/def2-TZVP} = \Delta_{\rm ST}^\text{PT3/def2-SVP} - \Delta_{\rm ST}^\text{PT2/def2-SVP} + \Delta_{\rm ST}^\text{PT2/def2-TZVP}$.
\item [b] Ref.~\citenum{Ghosh:2017gl}, tPBE on-top functional with the 6-31+G(d,p) basis set, WFP-3 GAS partition, UB3LYP/6-31G(d,p) geometries.
\item [c] Ref.~\citenum{Sharma2019}, tPBE on-top functional with the 6-31+G(d,p) basis set, DMRG $M = 500$, UB3LYP/6-31G(d,p) geometries from Ref.~\citenum{Ghosh:2017gl}.
\item [d] Ref.~\citenum{Beran2021}, DMRG $M = 1000$, $k \in [2,7]$: the 6-31G(d,p) basis set and UB3LYP/6-31G(d,p) geometries, $k \in [8,12]$: the cc-pVDZ basis set and PBE0/def2-SVP geometries.
\item [e] Ref.~\citenum{Luo:2018gn}, externally contracted MRCI with singles and doubles plus Davidson correction (ec-MRCISD+Q), the ANO-L-VTZP(C)/ANO-S-MB(H) basis set, DMRG $M = 1000$, UB3LYP/6-31G(d,p) geometries.
\item [f] Ref.~\citenum{Yang:2016by}, particle-particle random phase approximation (pp-RPA) with the B3LYP functional, the cc-pVDZ basis set, UB3LYP/6-31G$^\ast$ geometries.
\item [g] Ref.~\citenum{Hajgato2011}, focal point analysis up to coupled cluster theory with singles, doubles, and perturbative triples [CCSD(T)], B3LYP/cc-pVTZ geometries.
\end{tablenotes}

\end{threeparttable}
\end{table*}

Table~\ref{tab:acene_vert} reports the vertical $\Delta_{\rm ST}$ of oligoacenes at various levels of theories.
For a given basis set, all SA-DMRG-DSRG results show good agreements with each other.
The SA-DSRG-PT3 values deviate from those of SA-sq-LDSRG(2) by a maximum of 0.02~eV, while the SA-DSRG-PT2 data are underestimated by 0.04--0.11~eV in comparison to SA-DSRG-PT3.
In general, the vertical $\Delta_{\rm ST}$ of SA-DSRG decreases as the size of basis set increases.
For example, the SA-DSRG-PT3/def2-TZVP results are, on average, 0.04~eV lower than those of SA-DSRG-PT3/def2-SVP.
Overall, we observe a fairly constant deviation when varying the correlation hierarchy of SA-DSRG with a fixed basis set or vice versa, enabling the focal point analysis in SA-DSRG.\cite{East:1993cl,Li:2021sf}
The final SA-DSRG prediction for the vertical $\Delta_{\rm ST}$ of dodecacene is 0.22~eV, which is in excellent agreement with the DMRG-AC0 value of 0.24~eV.
For large acenes ($k \geq 8$), the CCSD(T) vertical $\Delta_{\rm ST}$ is overestimated in comparison to the SA-DSRG predictions, but GAS-PDFT and pp-RPA findings are typically underestimated.
Nonetheless, we find perfect agreement between the SA-DSRG values of naphthalene and the CC3/aug-cc-pVTZ value (3.17~eV).\cite{Veril2021}

\begin{table}[h!]
\begin{threeparttable}

\scriptsize
\renewcommand{\arraystretch}{1.15}
\caption{Errors of the DSRG-PT2/def2-SVP adiabatic $\Delta_{\rm ST}$ (in meV) for using the compression 3-RDM algorithm with a few selected $M'$ values.\tnote{a}}
\label{tab:acene_compress}

\begin{tabular*}{\columnwidth}{@{\extracolsep{\stretch{1}}} l *{4}{d{1.2}} d{1.3} @{}}
\hline
\hline
molecule & \multicolumn{1}{c}{1} & \multicolumn{1}{c}{1.5} & \multicolumn{1}{c}{2} & \multicolumn{1}{c}{3} & \multicolumn{1}{c}{exact} \\

\hline
hexacene & 0.58 & 0.10 & 0.02 & 0.00 & 0.602 \\
heptacene & 1.13 & 0.20 & 0.04 & 0.00 & 0.442 \\
octacene & 1.86 & 0.34 & 0.06 & 0.00 & 0.340 \\
nonacene & 2.72 & 0.51 & 0.08 & 0.01 & 0.282 \\
decacene & 3.83 & 0.74 & 0.11 & 0.01 & 0.253 \\
undecacene & 5.34 & 1.01 & 0.12 & 0.00 & 0.240 \\

\hline
\hline
\end{tabular*}

\begin{tablenotes}
\item [a] $M'$ values are given in multiples of $M = 600$. The last column lists the adiabatic $\Delta_{\rm ST}$ (in eV) computed using the exact 3-RDM.
\end{tablenotes}

\end{threeparttable}
\end{table}

We note that computing the vertical $\Delta_{\rm ST}$ only requires the 1- and 2-RDMs of the reference wave functions.
In contrast, 3-RDMs contribute to the DSRG adiabatic $\Delta_{\rm ST}$ and their construction leads to a memory bottleneck for the dodecacene computations with the (50,50) active space.
Therefore, we compute the DSRG energies of dodecacene using the MPS compression algorithm as explained in Sec.~\ref{sec:comb}.
The bond dimension of the intermediate state is set to $M' = 2M$.
This value of $M'$ is selected based on the benchmark for smaller acenes, as illustrated in Table~\ref{tab:acene_compress}.
For a given $M'$, the accuracy slightly deteriorates as moving to larger acenes, yet $M' = 1.5M$ is sufficient to achieve micro eV accuracy for undecacene.
This observation is consistent with that observed in the time-dependent NEVPT2 with MPS compression,\cite{Sokolov:2017gr} where $M' \approx 2M$ is found to be sufficiently accurate.

\begin{table*}[h!]
\begin{threeparttable}

\scriptsize
\renewcommand{\arraystretch}{1.15}
\caption{Adiabatic singlet-triplet splittings (in eV) of the $k$-acene ($k = 2, 3, \dots, 12$) series.}
\label{tab:acene_adia}

\begin{tabular*}{\textwidth}{@{\extracolsep{\stretch{1}}} c c *{11}{d{1.2}} @{}}
\hline
\hline
& & \multicolumn{5}{c}{SS-DMRG-DSRG\tnote{a}} \\
\cline{3-7}
& & \multicolumn{3}{c}{def2-SVP} & \multicolumn{2}{c}{def2-TZVP} & \multicolumn{1}{c}{ACI-} & \multicolumn{1}{c}{GAS-} & \multicolumn{3}{c}{DMRG-} \\
\cline{3-5} \cline{6-7} \cline{10-12}
$k$ & active space & \multicolumn{1}{c}{PT2} & \multicolumn{1}{c}{PT3} & \multicolumn{1}{c}{sq-L2} & \multicolumn{1}{c}{PT2} & \multicolumn{1}{c}{PT3}  & \multicolumn{1}{c}{DSRG-PT2\tnote{b}} & \multicolumn{1}{c}{PDFT\tnote{c}} & \multicolumn{1}{c}{PDFT\tnote{d}} & \multicolumn{1}{c}{AC0\tnote{e}} & \multicolumn{1}{c}{ec-MRCISD+Q\tnote{f}} & \multicolumn{1}{c}{CCSD(T)\tnote{g}} \\
\hline

2 & (10,10) & 2.63 & 2.85 & 2.85 & 2.63 & 2.88 & 2.66 & 2.81 & 2.91 & 2.96 & 2.72 & 2.85 \\
3 & (14,14) & 1.80 & 2.03 & 2.02 & 1.78 & 2.06 & 1.82 & 1.87 & 2.00 & 2.10 & 1.81 & 2.09 \\
4 & (18,18) & 1.23 & 1.46 & 1.45 & 1.21 & 1.48 & 1.21 & 1.25 & 1.37 & 1.50 & 1.23 & 1.45 \\
5 & (22,22) & 0.85 & 1.06 & 1.05 & 0.83 & 1.07 & 0.78 & 0.89 & 0.98 & 1.09 & 0.93 & 1.10 \\
6 & (26,26) & 0.60 & 0.78 & 0.77 & 0.57 & 0.79 & 0.49 & 0.65 & 0.73 & 0.79 & 0.67 & 0.77 \\
7 & (30,30) & 0.44 & 0.58 & 0.57 & 0.40 & 0.55 & 0.33 & 0.43 & 0.62 & 0.59 & 0.48 & 0.58 \\
8 & (34,34) & 0.34 & 0.44 & 0.43 & 0.29 & 0.41 &  & 0.28 &  &  & 0.33 & 0.40 \\
9 & (38,38) & 0.28 & 0.36 & 0.35 & 0.24 & 0.32 &  & 0.22 &  &  & 0.24 & 0.30 \\
10 & (42,42) & 0.25 & 0.31 &  & 0.21 & 0.26 &  & 0.22 &  &  &  & 0.20 \\
11 & (46,46) & 0.24 & 0.28 &  & 0.20 & [0.24] &  & 0.13 &  &  &  & 0.16 \\
12 & (50,50) & 0.24 & 0.27 &  & 0.20 & [0.23] &  &  &  &  &  &  \\

\hline
\hline
\end{tabular*}

\begin{tablenotes}
\item [a] This work, UB3LYP/6-31G(d,p) geometries from Ref.~\citenum{Ghosh:2017gl}. The DSRG-PT3 values in brackets are estimated using the additive assumption $\Delta_{\rm ST}^\text{PT3/def2-TZVP} = \Delta_{\rm ST}^\text{PT3/def2-SVP} - \Delta_{\rm ST}^\text{PT2/def2-SVP} + \Delta_{\rm ST}^\text{PT2/def2-TZVP}$.
\item [b] Ref.~\citenum{Schriber:2018hw}, DSRG $s = 0.5$~\sunit, the cc-pVTZ basis set, UB3LYP/6-31G(d) geometries, restricted Hartree-Fock orbitals.
\item [c] Ref.~\citenum{Ghosh:2017gl}, tPBE on-top functional with the 6-31+G(d,p) basis set, WFP-3 GAS partition, UB3LYP/6-31G(d,p) geometries.
\item [d] Ref.~\citenum{Sharma2019}, tPBE on-top functional with the 6-31+G(d,p) basis set, DMRG $M = 500$, UB3LYP/6-31G(d,p) geometries from Ref.~\citenum{Ghosh:2017gl}.
\item [e] Ref.~\citenum{Beran2021}, the 6-31G(d,p) basis set, DMRG $M = 1000$, UB3LYP/6-31G(d,p) geometries.
\item [f] Ref.~\citenum{Luo:2018gn}, the ANO-L-VTZP(C)/ANO-S-MB(H) basis set, DMRG $M = 1000$, UB3LYP/6-31G(d,p) geometries.
\item [g] Ref.~\citenum{Hajgato2011}, focal point analysis, B3LYP/cc-pVTZ geometries.
\end{tablenotes}

\end{threeparttable}
\end{table*}

\begin{figure}[h]
\centering
\includegraphics[width=0.6\columnwidth]{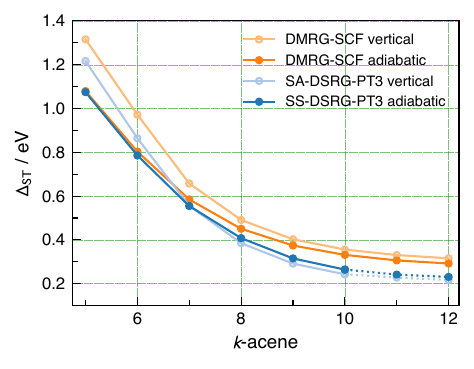}
\caption{Singlet--triplet gaps of oligoacenes obtained using DMRG-SCF and DSRG-PT3 with the def2-TZVP basis set. The DSRG-PT3 results of undecacene and dodecacene are estimated using the additive assumption $\Delta_{\rm ST}^\text{PT3/def2-TZVP} = \Delta_{\rm ST}^\text{PT3/def2-SVP} - \Delta_{\rm ST}^\text{PT2/def2-SVP} + \Delta_{\rm ST}^\text{PT2/def2-TZVP}$.}
\label{fig:acene}
\end{figure}

In Table~\ref{tab:acene_adia}, we list the adiabatic $\Delta_{\rm ST}$ of oligoacenes obtained using numerous theoretical methods.
Similar to the findings of vertical $\Delta_{\rm ST}$, the DSRG-PT3 and sq-LDSRG(2) methods yield nearly identical adiabatic $\Delta_{\rm ST}$ and they are higher than the DSRG-PT2 predictions by 0.15~eV on average with the def2-SVP basis set.
The DSRG adiabatic $\Delta_{\rm ST}$ is noticeably lower than the vertical counterpart for small acenes ($k < 8$).
As the acene size grows, the difference between the adiabatic and vertical $\Delta_{\rm ST}$ decreases to less than 0.04~eV (see Fig.~\ref{fig:acene}), indicating that the geometry plays a minor role in the $\Delta_{\rm ST}$ of large oligoacenes.
Indeed, previous findings also suggest highly analogous bond-length alternation pattern in singlet and triplet geometries in oligoacenes beyond octacene.\cite{Mullinax2019}
However, accurate predictions on adiabatic $\Delta_{\rm ST}$ may still require geometries that go beyond density function theory.
The comparison between the ACI- and DMRG-DSRG-PT2 data suggests that the ACI reference may be overly truncated for hexacene and heptacene as the DSRG-PT2 values of the two references differ by more than 0.07~eV.
Comparing to other methods, our best estimate of the adiabatic $\Delta_{\rm ST}$ from the DSRG-PT3/def2-TZVP method demonstrates surprising agreement with DMRG-AC0 and CCSD(T) with a mean absolute deviation of 0.03~eV.

\subsection{Vertical excitation energies of zeaxanthin}
\label{sec:zeaxanthin}

The SA-DMRG-DSRG vertical transitions of zeaxanthin are presented in Table~\ref{tab:zea_dsrg}.
In general, all SA-DSRG methods support the sequence of $\rm 2\, ^1 A_g^- < 1\, ^1 B_u^+ < 1\, ^1 B_u^-$.
Taking the SA-sq-LDSRG(2) data as reference, the predictions of SA-DSRG-PT2 are underestimated by 0.17~eV on average.
Increasing the size of basis set from double-$\zeta$ to triple-$\zeta$ significantly reduces the vertical transition energy (VTE) of $\rm 1\, ^1 B_u^+$ by 0.15~eV and 0.19~eV for SA-DSRG-PT2 and PT3, respectively.
The other two states are less sensitive to the basis set, resulting in a decrease in VTE from 0.05 to 0.07~eV.
In addition, Table~\ref{tab:zea_dsrg} includes the SA-DSRG-PT3/def2-TZVP results derived from the ground-state DMRG-SCF(22,22)/cc-pVDZ geometry.
Despite the consistent relative ordering of these states, the SA-DSRG-PT3/def2-TZVP VTEs obtained from different geometries show an average variation of 0.09~eV.

\begin{table*}[h!]
\begin{threeparttable}

\scriptsize
\renewcommand{\arraystretch}{1.15}
\caption{Vertical excitation energies ($\omega$, in eV) and oscillator strengths ($f$) of zeaxanthin obtained using various SA-DMRG-DSRG methods.}
\label{tab:zea_dsrg}

\begin{tabular*}{\textwidth}{@{\extracolsep{\stretch{1}}} c *{6}{d{1.2} d{1.3}} @{}}
\hline
\hline
& \multicolumn{6}{c}{def2-SV(P)} & \multicolumn{6}{c}{def2-TZVP} \\
\cline{2-7} \cline{8-13}
& \multicolumn{2}{c}{PT2} & \multicolumn{2}{c}{PT3} & \multicolumn{2}{c}{sq-L2} & \multicolumn{2}{c}{PT2} & \multicolumn{2}{c}{PT3} & \multicolumn{2}{c}{PT3\tnote{a}} \\
\cline{2-3} \cline{4-5} \cline{6-7} \cline{8-9} \cline{10-11} \cline{12-13}
state & \multicolumn{1}{c}{$\omega$} & \multicolumn{1}{c}{$f$} & \multicolumn{1}{c}{$\omega$} & \multicolumn{1}{c}{$f$} & \multicolumn{1}{c}{$\omega$} & \multicolumn{1}{c}{$f$} & \multicolumn{1}{c}{$\omega$} & \multicolumn{1}{c}{$f$} & \multicolumn{1}{c}{$\omega$} & \multicolumn{1}{c}{$f$} & \multicolumn{1}{c}{$\omega$} & \multicolumn{1}{c}{$f$} \\
\hline
$\rm 2\, ^1 A_{g}^{-}$ & 3.12 & 0.002 & 3.32 & 0.000 & 3.27 & 0.001 & 3.07 & 0.005 & 3.26 & 0.001 & 3.35 & 0.001 \\
$\rm 1\, ^1 B_{u}^{+}$ & 3.39 & 3.720 & 3.75 & 3.309 & 3.59 & 3.699 & 3.24 & 3.775 & 3.56 & 3.686 & 3.65 & 3.565 \\
$\rm 1\, ^1 B_{u}^{-}$ & 3.70 & 0.176 & 3.93 & 0.575 & 3.87 & 0.217 & 3.65 & 0.108 & 3.86 & 0.193 & 3.94 & 0.191 \\

\hline
\hline
\end{tabular*}

\begin{tablenotes}
\item [a] The ground-state DMRG-SCF(22,22)/cc-pVDZ geometry.
\end{tablenotes}

\end{threeparttable}
\end{table*}

In Table~\ref{tab:zea_vert}, we compare the VTEs of zeaxanthin obtained from SA-DMRG-DSRG and other theoretical approaches.
The effect of dynamical correlation is prominently evident in the optically bright $\rm 1\, ^1 B_u^+$ state, where the SA-DMRG VTE is lowered by over 1.85~eV as a result of the SA-DSRG treatment of electron correlation.
The current predictions of SA-DSRG-PT2 are generally higher than those previously reported using a larger flow parameter,\cite{Khokhlov2022} particularly for the $\rm 1\, ^1 B_u^+$ state.
However, the benchmark on butadiene reveals a substantial underestimation of the $\rm 1\, ^1 B_u^+$ state in SA-DSRG-PT2 using a converged flow parameter.\cite{Wang2023}
As also suggested by the def2-SV(P) results, we speculate that the SA-DSRG-PT3 VTEs of zeaxanthin are generally overestimated.
Indeed, the experimental value of the $\rm 1\, ^1 B_u^+$ state, albeit corresponding to adiabatic excitations, is only 2.48~eV.\cite{Josue2002}

\begin{table}[h!]
\begin{threeparttable}

\scriptsize
\renewcommand{\arraystretch}{1.15}
\caption{Theoretical estimates for the low-lying vertical excitation energies (in eV) of zeaxanthin.}
\label{tab:zea_vert}

\begin{tabular*}{\columnwidth}{@{\extracolsep{\stretch{1}}} c *{5}{d{1.2}} @{}}
\hline
\hline
& & & \multicolumn{3}{c}{SA-DMRG-DSRG} \\
\cline{4-6}
state & \multicolumn{1}{c}{DFT/MRCI\tnote{a}} & \multicolumn{1}{c}{SA-DMRG\tnote{b}} & \multicolumn{1}{c}{PT2\tnote{c}} & \multicolumn{1}{c}{PT2\tnote{b}} & \multicolumn{1}{c}{PT3\tnote{b}} \\

\hline
$\rm 2\, ^1  A_{g}^{-}$ & 2.11 & 3.59 & 2.92 & 3.07 & 3.26 \\
$\rm 1\, ^1  B_{u}^{+}$ & 2.48 & 5.41 & 2.90 & 3.24 & 3.56 \\
$\rm 1\, ^1  B_{u}^{-}$ & 2.75 & 4.19 &  & 3.65 & 3.86 \\

\hline
\hline
\end{tabular*}

\begin{tablenotes}
\item [a] Ref.~\citenum{Andreussi2015}, B3LYP/6-311G(d) geometry.
\item [b] This work, def2-TZVP basis set, CAM-B3LYP/cc-pVDZ geometry.
\item [c] Ref.~\citenum{Khokhlov2022}, cc-pVDZ basis set, CAM-B3LYP/cc-pVDZ geometry, flow parameter $s = 1.0$~\sunit.
\end{tablenotes}

\end{threeparttable}
\end{table}

\subsection{Ground-state potential energy curve of \ce{Cr2}}
\label{sec:cr2}

\begin{figure*}[h]
\centering
\includegraphics[width=0.95\textwidth]{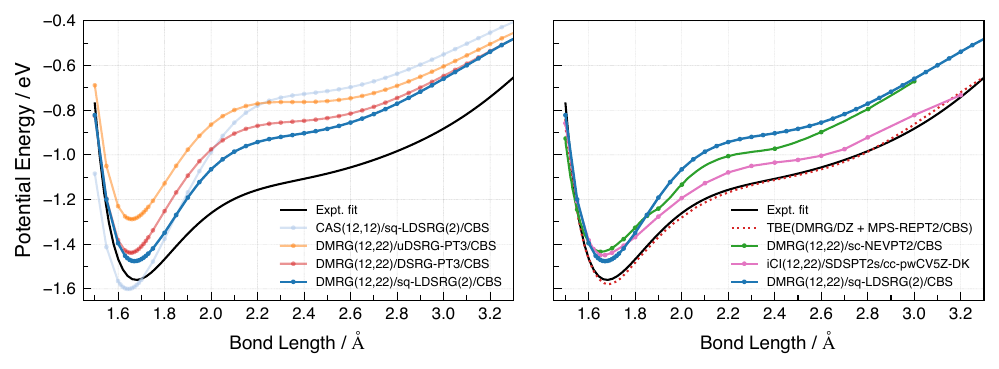}
\caption{The ground-state potential energy curves of \ce{Cr2}. The curves of theoretical best estimate (TBE) and experimental fit are taken from Ref.~\citenum{Larsson2022}.}
\label{fig:cr2}
\end{figure*}

Figure~\ref{fig:cr2} depicts the ground-state $X\, ^1\Sigma_{\rm g}^{+}$ potential energy curve (PEC) of \ce{Cr2}.
The experimental PEC fit\cite{Larsson2022} is shifted so that the bottom of the well meets the experimental dissociation energy of 1.56~eV.\cite{Simard1998}
As shown in the left panel of Fig.~\ref{fig:cr2}, the shape of the experimental PEC is qualitatively reproduced by DSRG-PT3 and sq-LDSRG(2).
However, relative to the dissociation limit, the DSRG energies are significantly underestimated by more than 0.16~eV for bond lengths ($r_{\rm Cr-Cr}$) greater than 1.9~\AA.
Analogous to previous findings,\cite{Kurashige:2011ck,Guo:2016fu,Vancoillie:2016gp} the use of double-shell active space leads to an improved description of the entire PEC over that of the valence active space.
Figure~\ref{fig:cr2} also presents the DSRG-PT3 results without reference relaxation [i.e., Eq.~\eqref{eq:dsrg_relax} is skipped], labeled as uDSRG-PT3.\cite{Li:2017bx}
Comparing to the default DSRG-PT3 that incorporates this effect, the reference relaxation deepens the potential well by 0.10--0.15~eV in the bonding region of 1.5--2.1~\AA.

In the right panel of Fig.~\ref{fig:cr2}, we compare the sq-LDSRG(2) PEC to a few selected theoretical methods, including the strongly contracted NEVPT2 (sc-NEVPT2)\cite{Guo:2016fu} and the static-dynamic-static second-order perturbation theory with selection (SDSPT2s) based on an iterative configuration interaction (iCI) reference.\cite{Lei2025}
These multireference perturbation theories yield comparable equilibrium bond distance and dissociation energy to those of sq-LDSRG(2).
However, the PEC of $r_{\rm Cr-Cr}$ > 1.9~\text{\AA} is better described by SDSPT2s than sc-NEVPT2 and sq-LDSRG(2).
This observation suggests that the coupling between excited configurations inside and outside the active space may be of importance for \ce{Cr2}.

We report the spectroscopic constants of the ground-state \ce{Cr2} in Table~\ref{tab:cr2}.
The DSRG values were obtained by fitting the PEC using 11 data points near the equilibrium based on the weighted least squares approach.\cite{Bender:2014do}
The DMRG-sq-LDSRG(2)/CBS predictions are in reasonable agreements with the theoretical best estimates (TBE),\cite{Larsson2022} deviated by 0.014~\AA, 10~\cm, and 0.10~eV for equilibrium bond length ($r_e$), harmonic vibrational frequency ($\omega_e$), and dissociation energy ($D_e$), respectively.
We also observe good agreements between DMRG-sq-LDSRG(2)/CBS and the experiment of Ref.~\citenum{Hilpert1987}, differing by 0.01~\AA, 26~\cm, and 0.01~eV for $r_e$, $\omega_e$, and $D_e$, respectively.

\begin{table*}[h!]
\begin{threeparttable}

\scriptsize
\renewcommand{\arraystretch}{1.15}
\caption{Spectroscopic constants (equilibrium bond length $r_e$, harmonic vibrational frequency $\omega_e$, fundamental vibrational frequency $\Delta G_{1/2}$, dissociation energy with respect to the bottom of the potential $D_e$) for the ground-state of \ce{Cr2}.}
\label{tab:cr2}

\begin{tabular*}{\textwidth}{@{\extracolsep{\stretch{1}}} l c l d{1.3} d{3.0} d{3.0} d{1.2} c @{}}
\hline
\hline
method & \multicolumn{1}{c}{active space} & basis set & \multicolumn{1}{c}{$r_{e}$ / \AA} & \multicolumn{1}{c}{$\omega_e$ / \cm} & \multicolumn{1}{c}{$\Delta G_{1/2}$ / \cm} & \multicolumn{1}{c}{$D_e$ / eV} & reference \\

\hline
sq-LDSRG(2) & (12,12) & CBS(QZ/5Z) & 1.644 & 572 & 556 & 1.60 & this work \\
DMRG-DSRG-PT3 & (12,22) & CBS(QZ/5Z) & 1.655 & 579 & 553 & 1.44 & this work \\
DMRG-sq-LDSRG(2) & (12,22) & CBS(QZ/5Z) & 1.671 & 505 & 487 & 1.48 & this work \\
DMRG-CASPT2/$\rm g_1$ & (12,28) & cc-pwCV5Z & 1.681 & 480 & & 1.61 & Ref.~\citenum{Kurashige:2011ck} \\
RASPT2\tnote{a} & (12,22) & CBS(QZ/5Z) & 1.666 & 489 & & 1.89 & Ref.~\citenum{Vancoillie:2016gp} \\
DMRG-sc-NEVPT2 & (12,22) & CBS(QZ/5Z) & 1.656 & 470 & & 1.43 & Ref.~\citenum{Guo:2016fu} \\
iCI-SDSPT2s & (12,22) & cc-pwCV5Z-DK & 1.659 & 462 & & 1.45 & Ref.~\citenum{Lei2025} \\
MRAQCC\tnote{b} & (12,12) & CBS(TZ/QZ) & 1.685 & 459 & 431 & 1.36 & Ref.~\citenum{Muller:2009jd} \\
DMRG-ec-MRCISD+Q & (12,42) & ANO-RCC-VQZP & 1.71 & 477 & & 1.62 & Ref.~\citenum{Luo:2018gn} \\
TBE & & CBS(QZ/5Z) & 1.685 & 495 & 468 & 1.58 & Ref.~\citenum{Larsson2022} \\
\hline
\multirow{4}{*}{experiments} & & & 1.6788 & 470 & \multicolumn{1}{c}{$452.34 \pm 0.02$} & & Ref.~\citenum{Bondybey1983} \\
& & & 1.68 & 479 & & \multicolumn{1}{c}{$1.472 \pm 0.052$} & Ref.~\citenum{Hilpert1987} \\
& & & & \multicolumn{1}{c}{$480.6 \pm 0.5$} & \multicolumn{1}{c}{$455 \pm 10$} & & Ref.~\citenum{Casey:1993gq} \\
& & & & & & \multicolumn{1}{c}{$1.56 \pm 0.06$\tnote{c}} & Ref.~\citenum{Simard1998} \\

\hline
\hline
\end{tabular*}

\begin{tablenotes}
\item [a] RASPT2: restricted active space second-order perturbation theory.
\item [b] MRAQCC: multireference averaged quadratic coupled cluster.
\item [c] A zero-point energy of 0.03~eV is applied to the experimental $D_0 = 1.53 \pm 0.06$ value.
\end{tablenotes}

\end{threeparttable}
\end{table*}

\section{Conclusions}
\label{sec:conclusion}

In this work, we have combined DMRG and SA-DSRG to achieve an accurate description of both static and dynamical electron correlation in strongly correlated molecular systems.
A straightforward joint of these two methods can be routinely applied to molecules requiring 30--40 active orbitals at either MRPT or MRCC level of the SA-DSRG theory.
Moreover, we have implemented the algorithm of forming the MPS-compressed intermediates,\cite{Sokolov:2017gr} which enables SA-DSRG-PT2 and PT3 computations on dodecacene with an active space of (50,50).
To the best of our knowledge, this active space is the largest employed in the internally contracted formalism of multireference theories.

We have performed DMRG-DSRG computations on the singlet-triplet gaps of oligoacenes, the low-lying vertical excitations of zeaxanthin, and the ground-state potential energy curve of chromium dimer.
For large oligoacenes, the vertical and adiabatic $\Delta_{\rm ST}$ values are found nearly identical, differing by less than 0.04~eV.
Our best estimates of the vertical and adiabatic $\Delta_{\rm ST}$ of dodecacene are 0.22 and 0.23~eV, respectively.
For zeaxanthin, we find that all SA-DSRG methods follow the order of $\rm 2\, ^1 A_g^- < 1\, ^1 B_u^+ < 1\, ^1 B_u^-$ on vertical excitation energies.
The shape of the \ce{Cr2} ground-state potential energy curve is also reasonably reproduced by DMRG(12,22)-sq-LDSRG(2), showing an equilibrium bond length of 1.671~\AA\ and an harmonic frequency of 505~\cm.

This work motivates further developments of DMRG-DSRG.
Given the previously noted inconsistencies of \ce{C-C} bond lengths in DFT and v2RDM optimized geometries,\cite{Mullinax2019} it may be useful to develop the analytic energy gradients\cite{Wang2021,Park2021c} for DMRG-DSRG-PT2 to assess the impact of molecular geometry on the oligoacene series.
Additionally, improving the DSRG-PT3 and LDSRG(2) implementations are essential for computations involving large basis sets.
Future benchmark studies on transition-metal complexes are also warranted for DMRG-based DSRG-PT3 and LDSRG(2) methods.
Nevertheless, we believe that DMRG-DSRG opens up new applications in the realm of conjugated molecules, including nanographenes,\cite{Yu2023a} chlorophylls, and carotenoids.\cite{Accomasso2024}

%

\section*{Acknowledgements}
This work was supported by the National Natural Science Foundation of China (Nos.~22103005, 22473013) and the Fundamental Research Funds for the Central Universities.
The Flatiron Institute is a division of the Simons Foundation.

\section*{Author Contributions}

\textbf{C.L.}: conceptualization, data curation, formal analysis, funding acquisition, investigation, methodology, resources, software, supervision, visualization, writing -- original draft, writing -- review \& editing.
\textbf{X.W.}: data curation, formal analysis, investigation, visualization, writing -- review \& editing.
\textbf{H.Z.}: formal analysis, methodology, software, writing -- review \& editing.
\textbf{W.F.}: funding acquisition, resources, writing -- review \& editing.


\begin{thebibliography}{116}%
\makeatletter
\providecommand \@ifxundefined [1]{%
 \@ifx{#1\undefined}
}%
\providecommand \@ifnum [1]{%
 \ifnum #1\expandafter \@firstoftwo
 \else \expandafter \@secondoftwo
 \fi
}%
\providecommand \@ifx [1]{%
 \ifx #1\expandafter \@firstoftwo
 \else \expandafter \@secondoftwo
 \fi
}%
\providecommand \natexlab [1]{#1}%
\providecommand \enquote  [1]{``#1''}%
\providecommand \bibnamefont  [1]{#1}%
\providecommand \bibfnamefont [1]{#1}%
\providecommand \citenamefont [1]{#1}%
\providecommand \href@noop [0]{\@secondoftwo}%
\providecommand \href [0]{\begingroup \@sanitize@url \@href}%
\providecommand \@href[1]{\@@startlink{#1}\@@href}%
\providecommand \@@href[1]{\endgroup#1\@@endlink}%
\providecommand \@sanitize@url [0]{\catcode `\\12\catcode `\$12\catcode
  `\&12\catcode `\#12\catcode `\^12\catcode `\_12\catcode `\%12\relax}%
\providecommand \@@startlink[1]{}%
\providecommand \@@endlink[0]{}%
\providecommand \url  [0]{\begingroup\@sanitize@url \@url }%
\providecommand \@url [1]{\endgroup\@href {#1}{\urlprefix }}%
\providecommand \urlprefix  [0]{URL }%
\providecommand \Eprint [0]{\href }%
\providecommand \doibase [0]{http://dx.doi.org/}%
\providecommand \selectlanguage [0]{\@gobble}%
\providecommand \bibinfo  [0]{\@secondoftwo}%
\providecommand \bibfield  [0]{\@secondoftwo}%
\providecommand \translation [1]{[#1]}%
\providecommand \BibitemOpen [0]{}%
\providecommand \bibitemStop [0]{}%
\providecommand \bibitemNoStop [0]{.\EOS\space}%
\providecommand \EOS [0]{\spacefactor3000\relax}%
\providecommand \BibitemShut  [1]{\csname bibitem#1\endcsname}%
\let\auto@bib@innerbib\@empty
\bibitem [{\citenamefont {Vogiatzis}\ \emph {et~al.}(2017)\citenamefont
  {Vogiatzis}, \citenamefont {Ma}, \citenamefont {Olsen}, \citenamefont
  {Gagliardi},\ and\ \citenamefont {de~Jong}}]{Vogiatzis:2017gp}%
  \BibitemOpen
  \bibfield  {author} {\bibinfo {author} {\bibfnamefont {K.~D.}\ \bibnamefont
  {Vogiatzis}}, \bibinfo {author} {\bibfnamefont {D.}~\bibnamefont {Ma}},
  \bibinfo {author} {\bibfnamefont {J.}~\bibnamefont {Olsen}}, \bibinfo
  {author} {\bibfnamefont {L.}~\bibnamefont {Gagliardi}}, \ and\ \bibinfo
  {author} {\bibfnamefont {W.~A.}\ \bibnamefont {de~Jong}},\ }\href {\doibase
  10.1063/1.4989858} {\bibfield  {journal} {\bibinfo  {journal} {J. Chem.
  Phys.}\ }\textbf {\bibinfo {volume} {147}},\ \bibinfo {pages} {184111}
  (\bibinfo {year} {2017})}\BibitemShut {NoStop}%
\bibitem [{\citenamefont {White}(1992)}]{White:1992ie}%
  \BibitemOpen
  \bibfield  {author} {\bibinfo {author} {\bibfnamefont {S.~R.}\ \bibnamefont
  {White}},\ }\href {\doibase 10.1103/PhysRevLett.69.2863} {\bibfield
  {journal} {\bibinfo  {journal} {Phys. Rev. Lett.}\ }\textbf {\bibinfo
  {volume} {69}},\ \bibinfo {pages} {2863} (\bibinfo {year}
  {1992})}\BibitemShut {NoStop}%
\bibitem [{\citenamefont {Schollw{\"{o}}ck}(2011)}]{Schollwock:2011gl}%
  \BibitemOpen
  \bibfield  {author} {\bibinfo {author} {\bibfnamefont {U.}~\bibnamefont
  {Schollw{\"{o}}ck}},\ }\href {\doibase 10.1016/j.aop.2010.09.012} {\bibfield
  {journal} {\bibinfo  {journal} {Ann. Phys. (N. Y).}\ }\textbf {\bibinfo
  {volume} {326}},\ \bibinfo {pages} {96} (\bibinfo {year} {2011})}\BibitemShut
  {NoStop}%
\bibitem [{\citenamefont {Chan}\ \emph {et~al.}(2016)\citenamefont {Chan},
  \citenamefont {Keselman}, \citenamefont {Nakatani}, \citenamefont {Li},\ and\
  \citenamefont {White}}]{Chan:2016fg}%
  \BibitemOpen
  \bibfield  {author} {\bibinfo {author} {\bibfnamefont {G.~K.-L.}\
  \bibnamefont {Chan}}, \bibinfo {author} {\bibfnamefont {A.}~\bibnamefont
  {Keselman}}, \bibinfo {author} {\bibfnamefont {N.}~\bibnamefont {Nakatani}},
  \bibinfo {author} {\bibfnamefont {Z.}~\bibnamefont {Li}}, \ and\ \bibinfo
  {author} {\bibfnamefont {S.~R.}\ \bibnamefont {White}},\ }\href {\doibase
  10.1063/1.4955108} {\bibfield  {journal} {\bibinfo  {journal} {J. Chem.
  Phys.}\ }\textbf {\bibinfo {volume} {145}},\ \bibinfo {pages} {014102}
  (\bibinfo {year} {2016})}\BibitemShut {NoStop}%
\bibitem [{\citenamefont {Evangelisti}, \citenamefont {Daudey},\ and\
  \citenamefont {Malrieu}(1983)}]{Evangelisti1983}%
  \BibitemOpen
  \bibfield  {author} {\bibinfo {author} {\bibfnamefont {S.}~\bibnamefont
  {Evangelisti}}, \bibinfo {author} {\bibfnamefont {J.-P.}\ \bibnamefont
  {Daudey}}, \ and\ \bibinfo {author} {\bibfnamefont {J.-P.}\ \bibnamefont
  {Malrieu}},\ }\href {\doibase 10.1016/0301-0104(83)85011-3} {\bibfield
  {journal} {\bibinfo  {journal} {Chem. Phys.}\ }\textbf {\bibinfo {volume}
  {75}},\ \bibinfo {pages} {91} (\bibinfo {year} {1983})}\BibitemShut {NoStop}%
\bibitem [{\citenamefont {Schriber}\ and\ \citenamefont
  {Evangelista}(2016)}]{Schriber:2016kl}%
  \BibitemOpen
  \bibfield  {author} {\bibinfo {author} {\bibfnamefont {J.~B.}\ \bibnamefont
  {Schriber}}\ and\ \bibinfo {author} {\bibfnamefont {F.~A.}\ \bibnamefont
  {Evangelista}},\ }\href {\doibase 10.1063/1.4948308} {\bibfield  {journal}
  {\bibinfo  {journal} {J. Chem. Phys.}\ }\textbf {\bibinfo {volume} {144}},\
  \bibinfo {pages} {161106} (\bibinfo {year} {2016})}\BibitemShut {NoStop}%
\bibitem [{\citenamefont {Holmes}, \citenamefont {Tubman},\ and\ \citenamefont
  {Umrigar}(2016)}]{Holmes:2016fm}%
  \BibitemOpen
  \bibfield  {author} {\bibinfo {author} {\bibfnamefont {A.~A.}\ \bibnamefont
  {Holmes}}, \bibinfo {author} {\bibfnamefont {N.~M.}\ \bibnamefont {Tubman}},
  \ and\ \bibinfo {author} {\bibfnamefont {C.~J.}\ \bibnamefont {Umrigar}},\
  }\href {\doibase 10.1021/acs.jctc.6b00407} {\bibfield  {journal} {\bibinfo
  {journal} {J. Chem. Theory Comput.}\ }\textbf {\bibinfo {volume} {12}},\
  \bibinfo {pages} {3674} (\bibinfo {year} {2016})}\BibitemShut {NoStop}%
\bibitem [{\citenamefont {Liu}\ and\ \citenamefont
  {Hoffmann}(2016)}]{Liu:2016jwa}%
  \BibitemOpen
  \bibfield  {author} {\bibinfo {author} {\bibfnamefont {W.}~\bibnamefont
  {Liu}}\ and\ \bibinfo {author} {\bibfnamefont {M.~R.}\ \bibnamefont
  {Hoffmann}},\ }\href {\doibase 10.1021/acs.jctc.5b01099} {\bibfield
  {journal} {\bibinfo  {journal} {J. Chem. Theory Comput.}\ }\textbf {\bibinfo
  {volume} {12}},\ \bibinfo {pages} {1169} (\bibinfo {year}
  {2016})}\BibitemShut {NoStop}%
\bibitem [{\citenamefont {Booth}, \citenamefont {Thom},\ and\ \citenamefont
  {Alavi}(2009)}]{Booth:2009hb}%
  \BibitemOpen
  \bibfield  {author} {\bibinfo {author} {\bibfnamefont {G.~H.}\ \bibnamefont
  {Booth}}, \bibinfo {author} {\bibfnamefont {A.~J.~W.}\ \bibnamefont {Thom}},
  \ and\ \bibinfo {author} {\bibfnamefont {A.}~\bibnamefont {Alavi}},\ }\href
  {\doibase 10.1063/1.3193710} {\bibfield  {journal} {\bibinfo  {journal} {J.
  Chem. Phys.}\ }\textbf {\bibinfo {volume} {131}},\ \bibinfo {pages} {054106}
  (\bibinfo {year} {2009})}\BibitemShut {NoStop}%
\bibitem [{\citenamefont {Gidofalvi}\ and\ \citenamefont
  {Mazziotti}(2008)}]{Gidofalvi2008}%
  \BibitemOpen
  \bibfield  {author} {\bibinfo {author} {\bibfnamefont {G.}~\bibnamefont
  {Gidofalvi}}\ and\ \bibinfo {author} {\bibfnamefont {D.~A.}\ \bibnamefont
  {Mazziotti}},\ }\href {\doibase 10.1063/1.2983652} {\bibfield  {journal}
  {\bibinfo  {journal} {J. Chem. Phys.}\ }\textbf {\bibinfo {volume} {129}},\
  \bibinfo {pages} {134108} (\bibinfo {year} {2008})}\BibitemShut {NoStop}%
\bibitem [{\citenamefont {Fosso-Tande}\ \emph {et~al.}(2016)\citenamefont
  {Fosso-Tande}, \citenamefont {Nguyen}, \citenamefont {Gidofalvi},\ and\
  \citenamefont {DePrince}}]{FossoTande:2016hb}%
  \BibitemOpen
  \bibfield  {author} {\bibinfo {author} {\bibfnamefont {J.}~\bibnamefont
  {Fosso-Tande}}, \bibinfo {author} {\bibfnamefont {T.-S.}\ \bibnamefont
  {Nguyen}}, \bibinfo {author} {\bibfnamefont {G.}~\bibnamefont {Gidofalvi}}, \
  and\ \bibinfo {author} {\bibfnamefont {A.~E.}\ \bibnamefont {DePrince}},\
  }\href {\doibase 10.1021/acs.jctc.6b00190} {\bibfield  {journal} {\bibinfo
  {journal} {J. Chem. Theory Comput.}\ }\textbf {\bibinfo {volume} {12}},\
  \bibinfo {pages} {2260} (\bibinfo {year} {2016})}\BibitemShut {NoStop}%
\bibitem [{\citenamefont {Sharma}\ \emph
  {et~al.}(2017{\natexlab{a}})\citenamefont {Sharma}, \citenamefont {Holmes},
  \citenamefont {Jeanmairet}, \citenamefont {Alavi},\ and\ \citenamefont
  {Umrigar}}]{Sharma:2017iu}%
  \BibitemOpen
  \bibfield  {author} {\bibinfo {author} {\bibfnamefont {S.}~\bibnamefont
  {Sharma}}, \bibinfo {author} {\bibfnamefont {A.~A.}\ \bibnamefont {Holmes}},
  \bibinfo {author} {\bibfnamefont {G.}~\bibnamefont {Jeanmairet}}, \bibinfo
  {author} {\bibfnamefont {A.}~\bibnamefont {Alavi}}, \ and\ \bibinfo {author}
  {\bibfnamefont {C.~J.}\ \bibnamefont {Umrigar}},\ }\href {\doibase
  10.1021/acs.jctc.6b01028} {\bibfield  {journal} {\bibinfo  {journal} {J.
  Chem. Theory Comput.}\ }\textbf {\bibinfo {volume} {13}},\ \bibinfo {pages}
  {1595} (\bibinfo {year} {2017}{\natexlab{a}})}\BibitemShut {NoStop}%
\bibitem [{\citenamefont {Zhai}\ and\ \citenamefont {Chan}(2021)}]{Zhai2021}%
  \BibitemOpen
  \bibfield  {author} {\bibinfo {author} {\bibfnamefont {H.}~\bibnamefont
  {Zhai}}\ and\ \bibinfo {author} {\bibfnamefont {G.~K.-L.}\ \bibnamefont
  {Chan}},\ }\href {\doibase 10.1063/5.0050902} {\bibfield  {journal} {\bibinfo
   {journal} {J. Chem. Phys.}\ }\textbf {\bibinfo {volume} {154}},\ \bibinfo
  {pages} {224116} (\bibinfo {year} {2021})},\ \Eprint
  {http://arxiv.org/abs/2103.09976} {arXiv:2103.09976} \BibitemShut {NoStop}%
\bibitem [{\citenamefont {Szalay}\ \emph {et~al.}(2012)\citenamefont {Szalay},
  \citenamefont {M{\"{u}}ller}, \citenamefont {Gidofalvi}, \citenamefont
  {Lischka},\ and\ \citenamefont {Shepard}}]{Szalay:2012df}%
  \BibitemOpen
  \bibfield  {author} {\bibinfo {author} {\bibfnamefont {P.~G.}\ \bibnamefont
  {Szalay}}, \bibinfo {author} {\bibfnamefont {T.}~\bibnamefont
  {M{\"{u}}ller}}, \bibinfo {author} {\bibfnamefont {G.}~\bibnamefont
  {Gidofalvi}}, \bibinfo {author} {\bibfnamefont {H.}~\bibnamefont {Lischka}},
  \ and\ \bibinfo {author} {\bibfnamefont {R.}~\bibnamefont {Shepard}},\ }\href
  {\doibase 10.1021/cr200137a} {\bibfield  {journal} {\bibinfo  {journal}
  {Chem. Rev.}\ }\textbf {\bibinfo {volume} {112}},\ \bibinfo {pages} {108}
  (\bibinfo {year} {2012})}\BibitemShut {NoStop}%
\bibitem [{\citenamefont {Luo}\ \emph {et~al.}(2018)\citenamefont {Luo},
  \citenamefont {Ma}, \citenamefont {Wang},\ and\ \citenamefont
  {Ma}}]{Luo:2018gn}%
  \BibitemOpen
  \bibfield  {author} {\bibinfo {author} {\bibfnamefont {Z.}~\bibnamefont
  {Luo}}, \bibinfo {author} {\bibfnamefont {Y.}~\bibnamefont {Ma}}, \bibinfo
  {author} {\bibfnamefont {X.}~\bibnamefont {Wang}}, \ and\ \bibinfo {author}
  {\bibfnamefont {H.}~\bibnamefont {Ma}},\ }\href {\doibase
  10.1021/acs.jctc.8b00613} {\bibfield  {journal} {\bibinfo  {journal} {J.
  Chem. Theory Comput.}\ }\textbf {\bibinfo {volume} {14}},\ \bibinfo {pages}
  {4747} (\bibinfo {year} {2018})}\BibitemShut {NoStop}%
\bibitem [{\citenamefont {Andersson}, \citenamefont {Malmqvist},\ and\
  \citenamefont {Roos}(1992)}]{Andersson:1992cq}%
  \BibitemOpen
  \bibfield  {author} {\bibinfo {author} {\bibfnamefont {K.}~\bibnamefont
  {Andersson}}, \bibinfo {author} {\bibfnamefont {P.-{\AA}.}\ \bibnamefont
  {Malmqvist}}, \ and\ \bibinfo {author} {\bibfnamefont {B.~O.}\ \bibnamefont
  {Roos}},\ }\href {\doibase 10.1063/1.462209} {\bibfield  {journal} {\bibinfo
  {journal} {J. Chem. Phys.}\ }\textbf {\bibinfo {volume} {96}},\ \bibinfo
  {pages} {1218} (\bibinfo {year} {1992})}\BibitemShut {NoStop}%
\bibitem [{\citenamefont {Angeli}\ \emph {et~al.}(2001)\citenamefont {Angeli},
  \citenamefont {Cimiraglia}, \citenamefont {Evangelisti}, \citenamefont
  {Leininger},\ and\ \citenamefont {Malrieu}}]{Angeli:2001bg}%
  \BibitemOpen
  \bibfield  {author} {\bibinfo {author} {\bibfnamefont {C.}~\bibnamefont
  {Angeli}}, \bibinfo {author} {\bibfnamefont {R.}~\bibnamefont {Cimiraglia}},
  \bibinfo {author} {\bibfnamefont {S.}~\bibnamefont {Evangelisti}}, \bibinfo
  {author} {\bibfnamefont {T.}~\bibnamefont {Leininger}}, \ and\ \bibinfo
  {author} {\bibfnamefont {J.-P.}\ \bibnamefont {Malrieu}},\ }\href {\doibase
  10.1063/1.1361246} {\bibfield  {journal} {\bibinfo  {journal} {J. Chem.
  Phys.}\ }\textbf {\bibinfo {volume} {114}},\ \bibinfo {pages} {10252}
  (\bibinfo {year} {2001})}\BibitemShut {NoStop}%
\bibitem [{\citenamefont {Sharma}\ \emph
  {et~al.}(2017{\natexlab{b}})\citenamefont {Sharma}, \citenamefont {Knizia},
  \citenamefont {Guo},\ and\ \citenamefont {Alavi}}]{Sharma:2017eo}%
  \BibitemOpen
  \bibfield  {author} {\bibinfo {author} {\bibfnamefont {S.}~\bibnamefont
  {Sharma}}, \bibinfo {author} {\bibfnamefont {G.}~\bibnamefont {Knizia}},
  \bibinfo {author} {\bibfnamefont {S.}~\bibnamefont {Guo}}, \ and\ \bibinfo
  {author} {\bibfnamefont {A.}~\bibnamefont {Alavi}},\ }\href {\doibase
  10.1021/acs.jctc.6b00898} {\bibfield  {journal} {\bibinfo  {journal} {J.
  Chem. Theory Comput.}\ }\textbf {\bibinfo {volume} {13}},\ \bibinfo {pages}
  {488} (\bibinfo {year} {2017}{\natexlab{b}})}\BibitemShut {NoStop}%
\bibitem [{\citenamefont {Lyakh}\ \emph {et~al.}(2012)\citenamefont {Lyakh},
  \citenamefont {Musia{\l}}, \citenamefont {Lotrich},\ and\ \citenamefont
  {Bartlett}}]{Lyakh:2012cn}%
  \BibitemOpen
  \bibfield  {author} {\bibinfo {author} {\bibfnamefont {D.~I.}\ \bibnamefont
  {Lyakh}}, \bibinfo {author} {\bibfnamefont {M.}~\bibnamefont {Musia{\l}}},
  \bibinfo {author} {\bibfnamefont {V.~F.}\ \bibnamefont {Lotrich}}, \ and\
  \bibinfo {author} {\bibfnamefont {R.~J.}\ \bibnamefont {Bartlett}},\ }\href
  {\doibase 10.1021/cr2001417} {\bibfield  {journal} {\bibinfo  {journal}
  {Chem. Rev.}\ }\textbf {\bibinfo {volume} {112}},\ \bibinfo {pages} {182}
  (\bibinfo {year} {2012})}\BibitemShut {NoStop}%
\bibitem [{\citenamefont {Evangelista}(2018)}]{Evangelista:2018bt}%
  \BibitemOpen
  \bibfield  {author} {\bibinfo {author} {\bibfnamefont {F.~A.}\ \bibnamefont
  {Evangelista}},\ }\href {\doibase 10.1063/1.5039496} {\bibfield  {journal}
  {\bibinfo  {journal} {J. Chem. Phys.}\ }\textbf {\bibinfo {volume} {149}},\
  \bibinfo {pages} {030901} (\bibinfo {year} {2018})}\BibitemShut {NoStop}%
\bibitem [{\citenamefont {Sokolov}\ \emph {et~al.}(2017)\citenamefont
  {Sokolov}, \citenamefont {Guo}, \citenamefont {Ronca},\ and\ \citenamefont
  {Chan}}]{Sokolov:2017gr}%
  \BibitemOpen
  \bibfield  {author} {\bibinfo {author} {\bibfnamefont {A.~Y.}\ \bibnamefont
  {Sokolov}}, \bibinfo {author} {\bibfnamefont {S.}~\bibnamefont {Guo}},
  \bibinfo {author} {\bibfnamefont {E.}~\bibnamefont {Ronca}}, \ and\ \bibinfo
  {author} {\bibfnamefont {G.~K.-L.}\ \bibnamefont {Chan}},\ }\href {\doibase
  10.1063/1.4986975} {\bibfield  {journal} {\bibinfo  {journal} {J. Chem.
  Phys.}\ }\textbf {\bibinfo {volume} {146}},\ \bibinfo {pages} {244102}
  (\bibinfo {year} {2017})}\BibitemShut {NoStop}%
\bibitem [{\citenamefont {Andersson}\ \emph {et~al.}(1994)\citenamefont
  {Andersson}, \citenamefont {Roos}, \citenamefont {Malmqvist},\ and\
  \citenamefont {Widmark}}]{Andersson1994}%
  \BibitemOpen
  \bibfield  {author} {\bibinfo {author} {\bibfnamefont {K.}~\bibnamefont
  {Andersson}}, \bibinfo {author} {\bibfnamefont {B.}~\bibnamefont {Roos}},
  \bibinfo {author} {\bibfnamefont {P.-{\AA}.}\ \bibnamefont {Malmqvist}}, \
  and\ \bibinfo {author} {\bibfnamefont {P.-O.}\ \bibnamefont {Widmark}},\
  }\href {\doibase 10.1016/0009-2614(94)01183-4} {\bibfield  {journal}
  {\bibinfo  {journal} {Chem. Phys. Lett.}\ }\textbf {\bibinfo {volume}
  {230}},\ \bibinfo {pages} {391} (\bibinfo {year} {1994})}\BibitemShut
  {NoStop}%
\bibitem [{\citenamefont {Larsson}\ \emph {et~al.}(2022)\citenamefont
  {Larsson}, \citenamefont {Zhai}, \citenamefont {Umrigar},\ and\ \citenamefont
  {Chan}}]{Larsson2022}%
  \BibitemOpen
  \bibfield  {author} {\bibinfo {author} {\bibfnamefont {H.~R.}\ \bibnamefont
  {Larsson}}, \bibinfo {author} {\bibfnamefont {H.}~\bibnamefont {Zhai}},
  \bibinfo {author} {\bibfnamefont {C.~J.}\ \bibnamefont {Umrigar}}, \ and\
  \bibinfo {author} {\bibfnamefont {G.~K.-L.}\ \bibnamefont {Chan}},\ }\href
  {\doibase 10.1021/jacs.2c06357} {\bibfield  {journal} {\bibinfo  {journal}
  {J. Am. Chem. Soc.}\ }\textbf {\bibinfo {volume} {144}},\ \bibinfo {pages}
  {15932} (\bibinfo {year} {2022})},\ \Eprint {http://arxiv.org/abs/2206.10738}
  {arXiv:2206.10738} \BibitemShut {NoStop}%
\bibitem [{\citenamefont {Kurashige}\ and\ \citenamefont
  {Yanai}(2011)}]{Kurashige:2011ck}%
  \BibitemOpen
  \bibfield  {author} {\bibinfo {author} {\bibfnamefont {Y.}~\bibnamefont
  {Kurashige}}\ and\ \bibinfo {author} {\bibfnamefont {T.}~\bibnamefont
  {Yanai}},\ }\href {\doibase 10.1063/1.3629454} {\bibfield  {journal}
  {\bibinfo  {journal} {J. Chem. Phys.}\ }\textbf {\bibinfo {volume} {135}},\
  \bibinfo {pages} {094104} (\bibinfo {year} {2011})}\BibitemShut {NoStop}%
\bibitem [{\citenamefont {Guo}\ \emph {et~al.}(2016)\citenamefont {Guo},
  \citenamefont {Watson}, \citenamefont {Hu}, \citenamefont {Sun},\ and\
  \citenamefont {Chan}}]{Guo:2016fu}%
  \BibitemOpen
  \bibfield  {author} {\bibinfo {author} {\bibfnamefont {S.}~\bibnamefont
  {Guo}}, \bibinfo {author} {\bibfnamefont {M.~A.}\ \bibnamefont {Watson}},
  \bibinfo {author} {\bibfnamefont {W.}~\bibnamefont {Hu}}, \bibinfo {author}
  {\bibfnamefont {Q.}~\bibnamefont {Sun}}, \ and\ \bibinfo {author}
  {\bibfnamefont {G.~K.-L.}\ \bibnamefont {Chan}},\ }\href {\doibase
  10.1021/acs.jctc.5b01225} {\bibfield  {journal} {\bibinfo  {journal} {J.
  Chem. Theory Comput.}\ }\textbf {\bibinfo {volume} {12}},\ \bibinfo {pages}
  {1583} (\bibinfo {year} {2016})}\BibitemShut {NoStop}%
\bibitem [{\citenamefont {Kurashige}\ \emph {et~al.}(2014)\citenamefont
  {Kurashige}, \citenamefont {Chalupsk{\'{y}}}, \citenamefont {Lan},\ and\
  \citenamefont {Yanai}}]{Kurashige:2014bq}%
  \BibitemOpen
  \bibfield  {author} {\bibinfo {author} {\bibfnamefont {Y.}~\bibnamefont
  {Kurashige}}, \bibinfo {author} {\bibfnamefont {J.}~\bibnamefont
  {Chalupsk{\'{y}}}}, \bibinfo {author} {\bibfnamefont {T.~N.}\ \bibnamefont
  {Lan}}, \ and\ \bibinfo {author} {\bibfnamefont {T.}~\bibnamefont {Yanai}},\
  }\href {\doibase 10.1063/1.4900878} {\bibfield  {journal} {\bibinfo
  {journal} {J. Chem. Phys.}\ }\textbf {\bibinfo {volume} {141}},\ \bibinfo
  {pages} {174111} (\bibinfo {year} {2014})}\BibitemShut {NoStop}%
\bibitem [{\citenamefont {Phung}, \citenamefont {Wouters},\ and\ \citenamefont
  {Pierloot}(2016)}]{Phung:2016ke}%
  \BibitemOpen
  \bibfield  {author} {\bibinfo {author} {\bibfnamefont {Q.~M.}\ \bibnamefont
  {Phung}}, \bibinfo {author} {\bibfnamefont {S.}~\bibnamefont {Wouters}}, \
  and\ \bibinfo {author} {\bibfnamefont {K.}~\bibnamefont {Pierloot}},\ }\href
  {\doibase 10.1021/acs.jctc.6b00714} {\bibfield  {journal} {\bibinfo
  {journal} {J. Chem. Theory Comput.}\ }\textbf {\bibinfo {volume} {12}},\
  \bibinfo {pages} {4352} (\bibinfo {year} {2016})}\BibitemShut {NoStop}%
\bibitem [{\citenamefont {{Li Manni}}\ \emph {et~al.}(2014)\citenamefont {{Li
  Manni}}, \citenamefont {Carlson}, \citenamefont {Luo}, \citenamefont {Ma},
  \citenamefont {Olsen}, \citenamefont {Truhlar},\ and\ \citenamefont
  {Gagliardi}}]{LiManni:2014kg}%
  \BibitemOpen
  \bibfield  {author} {\bibinfo {author} {\bibfnamefont {G.}~\bibnamefont {{Li
  Manni}}}, \bibinfo {author} {\bibfnamefont {R.~K.}\ \bibnamefont {Carlson}},
  \bibinfo {author} {\bibfnamefont {S.}~\bibnamefont {Luo}}, \bibinfo {author}
  {\bibfnamefont {D.}~\bibnamefont {Ma}}, \bibinfo {author} {\bibfnamefont
  {J.}~\bibnamefont {Olsen}}, \bibinfo {author} {\bibfnamefont {D.~G.}\
  \bibnamefont {Truhlar}}, \ and\ \bibinfo {author} {\bibfnamefont
  {L.}~\bibnamefont {Gagliardi}},\ }\href {\doibase 10.1021/ct500483t}
  {\bibfield  {journal} {\bibinfo  {journal} {J. Chem. Theory Comput.}\
  }\textbf {\bibinfo {volume} {10}},\ \bibinfo {pages} {3669} (\bibinfo {year}
  {2014})}\BibitemShut {NoStop}%
\bibitem [{\citenamefont {Pernal}(2018)}]{Pernal:2018fg}%
  \BibitemOpen
  \bibfield  {author} {\bibinfo {author} {\bibfnamefont {K.}~\bibnamefont
  {Pernal}},\ }\href {\doibase 10.1103/PhysRevLett.120.013001} {\bibfield
  {journal} {\bibinfo  {journal} {Phys. Rev. Lett.}\ }\textbf {\bibinfo
  {volume} {120}},\ \bibinfo {pages} {013001} (\bibinfo {year}
  {2018})}\BibitemShut {NoStop}%
\bibitem [{\citenamefont {Pastorczak}\ and\ \citenamefont
  {Pernal}(2018)}]{Pastorczak:2018fl}%
  \BibitemOpen
  \bibfield  {author} {\bibinfo {author} {\bibfnamefont {E.}~\bibnamefont
  {Pastorczak}}\ and\ \bibinfo {author} {\bibfnamefont {K.}~\bibnamefont
  {Pernal}},\ }\href {\doibase 10.1021/acs.jctc.8b00213} {\bibfield  {journal}
  {\bibinfo  {journal} {J. Chem. Theory Comput.}\ }\textbf {\bibinfo {volume}
  {14}},\ \bibinfo {pages} {3493} (\bibinfo {year} {2018})}\BibitemShut
  {NoStop}%
\bibitem [{\citenamefont {Ghosh}\ \emph {et~al.}(2017)\citenamefont {Ghosh},
  \citenamefont {Cramer}, \citenamefont {Truhlar},\ and\ \citenamefont
  {Gagliardi}}]{Ghosh:2017gl}%
  \BibitemOpen
  \bibfield  {author} {\bibinfo {author} {\bibfnamefont {S.}~\bibnamefont
  {Ghosh}}, \bibinfo {author} {\bibfnamefont {C.~J.}\ \bibnamefont {Cramer}},
  \bibinfo {author} {\bibfnamefont {D.~G.}\ \bibnamefont {Truhlar}}, \ and\
  \bibinfo {author} {\bibfnamefont {L.}~\bibnamefont {Gagliardi}},\ }\href
  {\doibase 10.1039/C6SC05036K} {\bibfield  {journal} {\bibinfo  {journal}
  {Chem. Sci.}\ }\textbf {\bibinfo {volume} {8}},\ \bibinfo {pages} {2741}
  (\bibinfo {year} {2017})}\BibitemShut {NoStop}%
\bibitem [{\citenamefont {Zuzak}\ \emph {et~al.}(2024)\citenamefont {Zuzak},
  \citenamefont {Kumar}, \citenamefont {Stoica}, \citenamefont {Soler‐Polo},
  \citenamefont {Brabec}, \citenamefont {Pernal}, \citenamefont {Veis},
  \citenamefont {Blieck}, \citenamefont {Echavarren}, \citenamefont {Jelinek},\
  and\ \citenamefont {Godlewski}}]{Zuzak2024}%
  \BibitemOpen
  \bibfield  {author} {\bibinfo {author} {\bibfnamefont {R.}~\bibnamefont
  {Zuzak}}, \bibinfo {author} {\bibfnamefont {M.}~\bibnamefont {Kumar}},
  \bibinfo {author} {\bibfnamefont {O.}~\bibnamefont {Stoica}}, \bibinfo
  {author} {\bibfnamefont {D.}~\bibnamefont {Soler‐Polo}}, \bibinfo {author}
  {\bibfnamefont {J.}~\bibnamefont {Brabec}}, \bibinfo {author} {\bibfnamefont
  {K.}~\bibnamefont {Pernal}}, \bibinfo {author} {\bibfnamefont
  {L.}~\bibnamefont {Veis}}, \bibinfo {author} {\bibfnamefont {R.}~\bibnamefont
  {Blieck}}, \bibinfo {author} {\bibfnamefont {A.~M.}\ \bibnamefont
  {Echavarren}}, \bibinfo {author} {\bibfnamefont {P.}~\bibnamefont {Jelinek}},
  \ and\ \bibinfo {author} {\bibfnamefont {S.}~\bibnamefont {Godlewski}},\
  }\href {\doibase 10.1002/anie.202317091} {\bibfield  {journal} {\bibinfo
  {journal} {Angew. Chemie Int. Ed.}\ }\textbf {\bibinfo {volume} {63}},\
  \bibinfo {pages} {e202317091} (\bibinfo {year} {2024})}\BibitemShut {NoStop}%
\bibitem [{\citenamefont {Evangelista}(2014)}]{Evangelista:2014kt}%
  \BibitemOpen
  \bibfield  {author} {\bibinfo {author} {\bibfnamefont {F.~A.}\ \bibnamefont
  {Evangelista}},\ }\href {\doibase 10.1063/1.4890660} {\bibfield  {journal}
  {\bibinfo  {journal} {J. Chem. Phys.}\ }\textbf {\bibinfo {volume} {141}},\
  \bibinfo {pages} {054109} (\bibinfo {year} {2014})}\BibitemShut {NoStop}%
\bibitem [{\citenamefont {Li}\ and\ \citenamefont
  {Evangelista}(2019)}]{Li:2019fu}%
  \BibitemOpen
  \bibfield  {author} {\bibinfo {author} {\bibfnamefont {C.}~\bibnamefont
  {Li}}\ and\ \bibinfo {author} {\bibfnamefont {F.~A.}\ \bibnamefont
  {Evangelista}},\ }\href {\doibase 10.1146/annurev-physchem-042018-052416}
  {\bibfield  {journal} {\bibinfo  {journal} {Annu. Rev. Phys. Chem.}\ }\textbf
  {\bibinfo {volume} {70}},\ \bibinfo {pages} {245} (\bibinfo {year}
  {2019})}\BibitemShut {NoStop}%
\bibitem [{\citenamefont {Hergert}\ \emph {et~al.}(2014)\citenamefont
  {Hergert}, \citenamefont {Bogner}, \citenamefont {Morris}, \citenamefont
  {Binder}, \citenamefont {Calci}, \citenamefont {Langhammer},\ and\
  \citenamefont {Roth}}]{Hergert:2014je}%
  \BibitemOpen
  \bibfield  {author} {\bibinfo {author} {\bibfnamefont {H.}~\bibnamefont
  {Hergert}}, \bibinfo {author} {\bibfnamefont {S.~K.}\ \bibnamefont {Bogner}},
  \bibinfo {author} {\bibfnamefont {T.~D.}\ \bibnamefont {Morris}}, \bibinfo
  {author} {\bibfnamefont {S.}~\bibnamefont {Binder}}, \bibinfo {author}
  {\bibfnamefont {A.}~\bibnamefont {Calci}}, \bibinfo {author} {\bibfnamefont
  {J.}~\bibnamefont {Langhammer}}, \ and\ \bibinfo {author} {\bibfnamefont
  {R.}~\bibnamefont {Roth}},\ }\href {\doibase 10.1103/PhysRevC.90.041302}
  {\bibfield  {journal} {\bibinfo  {journal} {Phys. Rev. C}\ }\textbf {\bibinfo
  {volume} {90}},\ \bibinfo {pages} {041302} (\bibinfo {year}
  {2014})}\BibitemShut {NoStop}%
\bibitem [{\citenamefont {Yanai}\ and\ \citenamefont
  {Chan}(2006)}]{Yanai:2006gi}%
  \BibitemOpen
  \bibfield  {author} {\bibinfo {author} {\bibfnamefont {T.}~\bibnamefont
  {Yanai}}\ and\ \bibinfo {author} {\bibfnamefont {G.~K.-L.}\ \bibnamefont
  {Chan}},\ }\href {\doibase 10.1063/1.2196410} {\bibfield  {journal} {\bibinfo
   {journal} {J. Chem. Phys.}\ }\textbf {\bibinfo {volume} {124}},\ \bibinfo
  {pages} {194106} (\bibinfo {year} {2006})}\BibitemShut {NoStop}%
\bibitem [{\citenamefont {Yanai}\ and\ \citenamefont
  {Chan}(2007)}]{Yanai:2007ix}%
  \BibitemOpen
  \bibfield  {author} {\bibinfo {author} {\bibfnamefont {T.}~\bibnamefont
  {Yanai}}\ and\ \bibinfo {author} {\bibfnamefont {G.~K.-L.}\ \bibnamefont
  {Chan}},\ }\href {\doibase 10.1063/1.2761870} {\bibfield  {journal} {\bibinfo
   {journal} {J. Chem. Phys.}\ }\textbf {\bibinfo {volume} {127}},\ \bibinfo
  {pages} {104107} (\bibinfo {year} {2007})}\BibitemShut {NoStop}%
\bibitem [{\citenamefont {Datta}, \citenamefont {Kong},\ and\ \citenamefont
  {Nooijen}(2011)}]{Datta:2011ca}%
  \BibitemOpen
  \bibfield  {author} {\bibinfo {author} {\bibfnamefont {D.}~\bibnamefont
  {Datta}}, \bibinfo {author} {\bibfnamefont {L.}~\bibnamefont {Kong}}, \ and\
  \bibinfo {author} {\bibfnamefont {M.}~\bibnamefont {Nooijen}},\ }\href
  {\doibase 10.1063/1.3592494} {\bibfield  {journal} {\bibinfo  {journal} {J.
  Chem. Phys.}\ }\textbf {\bibinfo {volume} {134}},\ \bibinfo {pages} {214116}
  (\bibinfo {year} {2011})}\BibitemShut {NoStop}%
\bibitem [{\citenamefont {Hanauer}\ and\ \citenamefont
  {K{\"{o}}hn}(2011)}]{Hanauer:2011ey}%
  \BibitemOpen
  \bibfield  {author} {\bibinfo {author} {\bibfnamefont {M.}~\bibnamefont
  {Hanauer}}\ and\ \bibinfo {author} {\bibfnamefont {A.}~\bibnamefont
  {K{\"{o}}hn}},\ }\href {\doibase 10.1063/1.3592786} {\bibfield  {journal}
  {\bibinfo  {journal} {J. Chem. Phys.}\ }\textbf {\bibinfo {volume} {134}},\
  \bibinfo {pages} {204111} (\bibinfo {year} {2011})}\BibitemShut {NoStop}%
\bibitem [{\citenamefont {Li}\ and\ \citenamefont
  {Evangelista}(2016)}]{Li:2016hb}%
  \BibitemOpen
  \bibfield  {author} {\bibinfo {author} {\bibfnamefont {C.}~\bibnamefont
  {Li}}\ and\ \bibinfo {author} {\bibfnamefont {F.~A.}\ \bibnamefont
  {Evangelista}},\ }\href {\doibase 10.1063/1.4947218} {\bibfield  {journal}
  {\bibinfo  {journal} {J. Chem. Phys.}\ }\textbf {\bibinfo {volume} {144}},\
  \bibinfo {pages} {164114} (\bibinfo {year} {2016})}\BibitemShut {NoStop}%
\bibitem [{\citenamefont {Li}\ and\ \citenamefont
  {Evangelista}(2015)}]{Li:2015iz}%
  \BibitemOpen
  \bibfield  {author} {\bibinfo {author} {\bibfnamefont {C.}~\bibnamefont
  {Li}}\ and\ \bibinfo {author} {\bibfnamefont {F.~A.}\ \bibnamefont
  {Evangelista}},\ }\href {\doibase 10.1021/acs.jctc.5b00134} {\bibfield
  {journal} {\bibinfo  {journal} {J. Chem. Theory Comput.}\ }\textbf {\bibinfo
  {volume} {11}},\ \bibinfo {pages} {2097} (\bibinfo {year}
  {2015})}\BibitemShut {NoStop}%
\bibitem [{\citenamefont {Li}\ and\ \citenamefont
  {Evangelista}(2017)}]{Li:2017bx}%
  \BibitemOpen
  \bibfield  {author} {\bibinfo {author} {\bibfnamefont {C.}~\bibnamefont
  {Li}}\ and\ \bibinfo {author} {\bibfnamefont {F.~A.}\ \bibnamefont
  {Evangelista}},\ }\href {\doibase 10.1063/1.4979016} {\bibfield  {journal}
  {\bibinfo  {journal} {J. Chem. Phys.}\ }\textbf {\bibinfo {volume} {146}},\
  \bibinfo {pages} {124132} (\bibinfo {year} {2017})}\BibitemShut {NoStop}%
\bibitem [{\citenamefont {Li}\ and\ \citenamefont
  {Evangelista}(2018)}]{Li:2018kl}%
  \BibitemOpen
  \bibfield  {author} {\bibinfo {author} {\bibfnamefont {C.}~\bibnamefont
  {Li}}\ and\ \bibinfo {author} {\bibfnamefont {F.~A.}\ \bibnamefont
  {Evangelista}},\ }\href {\doibase 10.1063/1.5019793} {\bibfield  {journal}
  {\bibinfo  {journal} {J. Chem. Phys.}\ }\textbf {\bibinfo {volume} {148}},\
  \bibinfo {pages} {124106} (\bibinfo {year} {2018})}\BibitemShut {NoStop}%
\bibitem [{\citenamefont {Zhang}, \citenamefont {Li},\ and\ \citenamefont
  {Evangelista}(2019)}]{Zhang:2019wj}%
  \BibitemOpen
  \bibfield  {author} {\bibinfo {author} {\bibfnamefont {T.}~\bibnamefont
  {Zhang}}, \bibinfo {author} {\bibfnamefont {C.}~\bibnamefont {Li}}, \ and\
  \bibinfo {author} {\bibfnamefont {F.~A.}\ \bibnamefont {Evangelista}},\
  }\href {\doibase 10.1021/acs.jctc.9b00353} {\bibfield  {journal} {\bibinfo
  {journal} {J. Chem. Theory Comput.}\ }\textbf {\bibinfo {volume} {15}},\
  \bibinfo {pages} {4399} (\bibinfo {year} {2019})}\BibitemShut {NoStop}%
\bibitem [{\citenamefont {Li}\ and\ \citenamefont
  {Evangelista}(2021)}]{Li:2021sf}%
  \BibitemOpen
  \bibfield  {author} {\bibinfo {author} {\bibfnamefont {C.}~\bibnamefont
  {Li}}\ and\ \bibinfo {author} {\bibfnamefont {F.~A.}\ \bibnamefont
  {Evangelista}},\ }\href {\doibase 10.1063/5.0059362} {\bibfield  {journal}
  {\bibinfo  {journal} {J. Chem. Phys.}\ }\textbf {\bibinfo {volume} {155}},\
  \bibinfo {pages} {114111} (\bibinfo {year} {2021})},\ \Eprint
  {http://arxiv.org/abs/2106.07097} {arXiv:2106.07097} \BibitemShut {NoStop}%
\bibitem [{\citenamefont {Phung}, \citenamefont {Nam},\ and\ \citenamefont
  {Saitow}(2023)}]{Phung2023}%
  \BibitemOpen
  \bibfield  {author} {\bibinfo {author} {\bibfnamefont {Q.~M.}\ \bibnamefont
  {Phung}}, \bibinfo {author} {\bibfnamefont {H.~N.}\ \bibnamefont {Nam}}, \
  and\ \bibinfo {author} {\bibfnamefont {M.}~\bibnamefont {Saitow}},\ }\href
  {\doibase 10.1021/acs.jpca.3c04254} {\bibfield  {journal} {\bibinfo
  {journal} {J. Phys. Chem. A}\ }\textbf {\bibinfo {volume} {127}},\ \bibinfo
  {pages} {7544} (\bibinfo {year} {2023})}\BibitemShut {NoStop}%
\bibitem [{\citenamefont {Wang}, \citenamefont {Fang},\ and\ \citenamefont
  {Li}(2023)}]{Wang2023}%
  \BibitemOpen
  \bibfield  {author} {\bibinfo {author} {\bibfnamefont {M.}~\bibnamefont
  {Wang}}, \bibinfo {author} {\bibfnamefont {W.-H.}\ \bibnamefont {Fang}}, \
  and\ \bibinfo {author} {\bibfnamefont {C.}~\bibnamefont {Li}},\ }\href
  {\doibase 10.1021/acs.jctc.2c00966} {\bibfield  {journal} {\bibinfo
  {journal} {J. Chem. Theory Comput.}\ }\textbf {\bibinfo {volume} {19}},\
  \bibinfo {pages} {122} (\bibinfo {year} {2023})}\BibitemShut {NoStop}%
\bibitem [{\citenamefont {Huang}\ and\ \citenamefont
  {Evangelista}(2023)}]{Huang2022d}%
  \BibitemOpen
  \bibfield  {author} {\bibinfo {author} {\bibfnamefont {M.}~\bibnamefont
  {Huang}}\ and\ \bibinfo {author} {\bibfnamefont {F.~A.}\ \bibnamefont
  {Evangelista}},\ }\href {\doibase 10.1063/5.0137096} {\bibfield  {journal}
  {\bibinfo  {journal} {J. Chem. Phys.}\ }\textbf {\bibinfo {volume} {158}},\
  \bibinfo {pages} {124112} (\bibinfo {year} {2023})},\ \Eprint
  {http://arxiv.org/abs/2212.04369} {arXiv:2212.04369} \BibitemShut {NoStop}%
\bibitem [{\citenamefont {Huang}\ and\ \citenamefont
  {Evangelista}(2024)}]{Huang2024b}%
  \BibitemOpen
  \bibfield  {author} {\bibinfo {author} {\bibfnamefont {M.}~\bibnamefont
  {Huang}}\ and\ \bibinfo {author} {\bibfnamefont {F.~A.}\ \bibnamefont
  {Evangelista}},\ }\href {\doibase 10.1021/acs.jctc.4c00835} {\bibfield
  {journal} {\bibinfo  {journal} {J. Chem. Theory Comput.}\ }\textbf {\bibinfo
  {volume} {20}},\ \bibinfo {pages} {7990} (\bibinfo {year}
  {2024})}\BibitemShut {NoStop}%
\bibitem [{\citenamefont {Schriber}\ \emph {et~al.}(2018)\citenamefont
  {Schriber}, \citenamefont {Hannon}, \citenamefont {Li},\ and\ \citenamefont
  {Evangelista}}]{Schriber:2018hw}%
  \BibitemOpen
  \bibfield  {author} {\bibinfo {author} {\bibfnamefont {J.~B.}\ \bibnamefont
  {Schriber}}, \bibinfo {author} {\bibfnamefont {K.~P.}\ \bibnamefont
  {Hannon}}, \bibinfo {author} {\bibfnamefont {C.}~\bibnamefont {Li}}, \ and\
  \bibinfo {author} {\bibfnamefont {F.~A.}\ \bibnamefont {Evangelista}},\
  }\href {\doibase 10.1021/acs.jctc.8b00877} {\bibfield  {journal} {\bibinfo
  {journal} {J. Chem. Theory Comput.}\ }\textbf {\bibinfo {volume} {14}},\
  \bibinfo {pages} {6295} (\bibinfo {year} {2018})}\BibitemShut {NoStop}%
\bibitem [{\citenamefont {Khokhlov}\ and\ \citenamefont
  {Belov}(2021)}]{Khokhlov2021}%
  \BibitemOpen
  \bibfield  {author} {\bibinfo {author} {\bibfnamefont {D.}~\bibnamefont
  {Khokhlov}}\ and\ \bibinfo {author} {\bibfnamefont {A.}~\bibnamefont
  {Belov}},\ }\href {\doibase 10.1021/acs.jctc.0c01293} {\bibfield  {journal}
  {\bibinfo  {journal} {J. Chem. Theory Comput.}\ }\textbf {\bibinfo {volume}
  {17}},\ \bibinfo {pages} {4301} (\bibinfo {year} {2021})}\BibitemShut
  {NoStop}%
\bibitem [{\citenamefont {Khokhlov}\ and\ \citenamefont
  {Belov}(2022)}]{Khokhlov2022}%
  \BibitemOpen
  \bibfield  {author} {\bibinfo {author} {\bibfnamefont {D.}~\bibnamefont
  {Khokhlov}}\ and\ \bibinfo {author} {\bibfnamefont {A.}~\bibnamefont
  {Belov}},\ }\href {\doibase 10.1021/acs.jpca.2c02485} {\bibfield  {journal}
  {\bibinfo  {journal} {J. Phys. Chem. A}\ }\textbf {\bibinfo {volume} {126}},\
  \bibinfo {pages} {4376} (\bibinfo {year} {2022})}\BibitemShut {NoStop}%
\bibitem [{\citenamefont {Li}\ \emph {et~al.}(2017)\citenamefont {Li},
  \citenamefont {Verma}, \citenamefont {Hannon},\ and\ \citenamefont
  {Evangelista}}]{Li:2017ff}%
  \BibitemOpen
  \bibfield  {author} {\bibinfo {author} {\bibfnamefont {C.}~\bibnamefont
  {Li}}, \bibinfo {author} {\bibfnamefont {P.}~\bibnamefont {Verma}}, \bibinfo
  {author} {\bibfnamefont {K.~P.}\ \bibnamefont {Hannon}}, \ and\ \bibinfo
  {author} {\bibfnamefont {F.~A.}\ \bibnamefont {Evangelista}},\ }\href
  {\doibase 10.1063/1.4997480} {\bibfield  {journal} {\bibinfo  {journal} {J.
  Chem. Phys.}\ }\textbf {\bibinfo {volume} {147}},\ \bibinfo {pages} {074107}
  (\bibinfo {year} {2017})}\BibitemShut {NoStop}%
\bibitem [{\citenamefont {T{\"{o}}nshoff}\ and\ \citenamefont
  {Bettinger}(2021)}]{Tonshoff2021}%
  \BibitemOpen
  \bibfield  {author} {\bibinfo {author} {\bibfnamefont {C.}~\bibnamefont
  {T{\"{o}}nshoff}}\ and\ \bibinfo {author} {\bibfnamefont {H.~F.}\
  \bibnamefont {Bettinger}},\ }\href {\doibase 10.1002/chem.202003112}
  {\bibfield  {journal} {\bibinfo  {journal} {Chem. – A Eur. J.}\ }\textbf
  {\bibinfo {volume} {27}},\ \bibinfo {pages} {3193} (\bibinfo {year}
  {2021})}\BibitemShut {NoStop}%
\bibitem [{\citenamefont {Hachmann}\ \emph {et~al.}(2007)\citenamefont
  {Hachmann}, \citenamefont {Dorando}, \citenamefont {Avil{\'{e}}s},\ and\
  \citenamefont {Chan}}]{Hachmann:2007ft}%
  \BibitemOpen
  \bibfield  {author} {\bibinfo {author} {\bibfnamefont {J.}~\bibnamefont
  {Hachmann}}, \bibinfo {author} {\bibfnamefont {J.~J.}\ \bibnamefont
  {Dorando}}, \bibinfo {author} {\bibfnamefont {M.}~\bibnamefont
  {Avil{\'{e}}s}}, \ and\ \bibinfo {author} {\bibfnamefont {G.~K.-L.}\
  \bibnamefont {Chan}},\ }\href {\doibase 10.1063/1.2768362} {\bibfield
  {journal} {\bibinfo  {journal} {J. Chem. Phys.}\ }\textbf {\bibinfo {volume}
  {127}},\ \bibinfo {pages} {134309} (\bibinfo {year} {2007})}\BibitemShut
  {NoStop}%
\bibitem [{\citenamefont {Yang}, \citenamefont {Davidson},\ and\ \citenamefont
  {Yang}(2016)}]{Yang:2016by}%
  \BibitemOpen
  \bibfield  {author} {\bibinfo {author} {\bibfnamefont {Y.}~\bibnamefont
  {Yang}}, \bibinfo {author} {\bibfnamefont {E.~R.}\ \bibnamefont {Davidson}},
  \ and\ \bibinfo {author} {\bibfnamefont {W.}~\bibnamefont {Yang}},\ }\href
  {\doibase 10.1073/pnas.1606021113} {\bibfield  {journal} {\bibinfo  {journal}
  {Proc. Natl. Acad. Sci.}\ }\textbf {\bibinfo {volume} {113}},\ \bibinfo
  {pages} {E5098} (\bibinfo {year} {2016})}\BibitemShut {NoStop}%
\bibitem [{\citenamefont {Trinquier}, \citenamefont {David},\ and\
  \citenamefont {Malrieu}(2018)}]{Trinquier2018}%
  \BibitemOpen
  \bibfield  {author} {\bibinfo {author} {\bibfnamefont {G.}~\bibnamefont
  {Trinquier}}, \bibinfo {author} {\bibfnamefont {G.}~\bibnamefont {David}}, \
  and\ \bibinfo {author} {\bibfnamefont {J.-P.}\ \bibnamefont {Malrieu}},\
  }\href {\doibase 10.1021/acs.jpca.8b03344} {\bibfield  {journal} {\bibinfo
  {journal} {J. Phys. Chem. A}\ }\textbf {\bibinfo {volume} {122}},\ \bibinfo
  {pages} {6926} (\bibinfo {year} {2018})}\BibitemShut {NoStop}%
\bibitem [{\citenamefont {Mondal}, \citenamefont {Shah},\ and\ \citenamefont
  {Neckers}(2006)}]{Mondai2006}%
  \BibitemOpen
  \bibfield  {author} {\bibinfo {author} {\bibfnamefont {R.}~\bibnamefont
  {Mondal}}, \bibinfo {author} {\bibfnamefont {B.~K.}\ \bibnamefont {Shah}}, \
  and\ \bibinfo {author} {\bibfnamefont {D.~C.}\ \bibnamefont {Neckers}},\
  }\href {\doibase 10.1021/ja063823i} {\bibfield  {journal} {\bibinfo
  {journal} {J. Am. Chem. Soc.}\ }\textbf {\bibinfo {volume} {128}},\ \bibinfo
  {pages} {9612} (\bibinfo {year} {2006})}\BibitemShut {NoStop}%
\bibitem [{\citenamefont {Shen}\ \emph {et~al.}(2018)\citenamefont {Shen},
  \citenamefont {Tatchen}, \citenamefont {Sanchez‐Garcia},\ and\
  \citenamefont {Bettinger}}]{Shen2018}%
  \BibitemOpen
  \bibfield  {author} {\bibinfo {author} {\bibfnamefont {B.}~\bibnamefont
  {Shen}}, \bibinfo {author} {\bibfnamefont {J.}~\bibnamefont {Tatchen}},
  \bibinfo {author} {\bibfnamefont {E.}~\bibnamefont {Sanchez‐Garcia}}, \
  and\ \bibinfo {author} {\bibfnamefont {H.~F.}\ \bibnamefont {Bettinger}},\
  }\href {\doibase 10.1002/anie.201802197} {\bibfield  {journal} {\bibinfo
  {journal} {Angew. Chemie Int. Ed.}\ }\textbf {\bibinfo {volume} {57}},\
  \bibinfo {pages} {10506} (\bibinfo {year} {2018})}\BibitemShut {NoStop}%
\bibitem [{\citenamefont {Zuzak}\ \emph {et~al.}(2018)\citenamefont {Zuzak},
  \citenamefont {Dorel}, \citenamefont {Kolmer}, \citenamefont {Szymonski},
  \citenamefont {Godlewski},\ and\ \citenamefont {Echavarren}}]{Zuzak2018}%
  \BibitemOpen
  \bibfield  {author} {\bibinfo {author} {\bibfnamefont {R.}~\bibnamefont
  {Zuzak}}, \bibinfo {author} {\bibfnamefont {R.}~\bibnamefont {Dorel}},
  \bibinfo {author} {\bibfnamefont {M.}~\bibnamefont {Kolmer}}, \bibinfo
  {author} {\bibfnamefont {M.}~\bibnamefont {Szymonski}}, \bibinfo {author}
  {\bibfnamefont {S.}~\bibnamefont {Godlewski}}, \ and\ \bibinfo {author}
  {\bibfnamefont {A.~M.}\ \bibnamefont {Echavarren}},\ }\href {\doibase
  10.1002/anie.201802040} {\bibfield  {journal} {\bibinfo  {journal} {Angew.
  Chemie Int. Ed.}\ }\textbf {\bibinfo {volume} {57}},\ \bibinfo {pages}
  {10500} (\bibinfo {year} {2018})}\BibitemShut {NoStop}%
\bibitem [{\citenamefont {Eisenhut}\ \emph {et~al.}(2020)\citenamefont
  {Eisenhut}, \citenamefont {K{\"{u}}hne}, \citenamefont {Garc{\'{i}}a},
  \citenamefont {Fern{\'{a}}ndez}, \citenamefont {Guiti{\'{a}}n}, \citenamefont
  {P{\'{e}}rez}, \citenamefont {Trinquier}, \citenamefont {Cuniberti},
  \citenamefont {Joachim}, \citenamefont {Pe{\~{n}}a},\ and\ \citenamefont
  {Moresco}}]{Eisenhut2020}%
  \BibitemOpen
  \bibfield  {author} {\bibinfo {author} {\bibfnamefont {F.}~\bibnamefont
  {Eisenhut}}, \bibinfo {author} {\bibfnamefont {T.}~\bibnamefont
  {K{\"{u}}hne}}, \bibinfo {author} {\bibfnamefont {F.}~\bibnamefont
  {Garc{\'{i}}a}}, \bibinfo {author} {\bibfnamefont {S.}~\bibnamefont
  {Fern{\'{a}}ndez}}, \bibinfo {author} {\bibfnamefont {E.}~\bibnamefont
  {Guiti{\'{a}}n}}, \bibinfo {author} {\bibfnamefont {D.}~\bibnamefont
  {P{\'{e}}rez}}, \bibinfo {author} {\bibfnamefont {G.}~\bibnamefont
  {Trinquier}}, \bibinfo {author} {\bibfnamefont {G.}~\bibnamefont
  {Cuniberti}}, \bibinfo {author} {\bibfnamefont {C.}~\bibnamefont {Joachim}},
  \bibinfo {author} {\bibfnamefont {D.}~\bibnamefont {Pe{\~{n}}a}}, \ and\
  \bibinfo {author} {\bibfnamefont {F.}~\bibnamefont {Moresco}},\ }\href
  {\doibase 10.1021/acsnano.9b08456} {\bibfield  {journal} {\bibinfo  {journal}
  {ACS Nano}\ }\textbf {\bibinfo {volume} {14}},\ \bibinfo {pages} {1011}
  (\bibinfo {year} {2020})}\BibitemShut {NoStop}%
\bibitem [{\citenamefont {Ruan}\ \emph {et~al.}(2025)\citenamefont {Ruan},
  \citenamefont {Schramm}, \citenamefont {Bauer}, \citenamefont {Naumann},
  \citenamefont {M{\"{u}}ller}, \citenamefont {S{\"{a}}ttele}, \citenamefont
  {Bettinger}, \citenamefont {Tonner-Zech},\ and\ \citenamefont
  {Gottfried}}]{Ruan2025}%
  \BibitemOpen
  \bibfield  {author} {\bibinfo {author} {\bibfnamefont {Z.}~\bibnamefont
  {Ruan}}, \bibinfo {author} {\bibfnamefont {J.}~\bibnamefont {Schramm}},
  \bibinfo {author} {\bibfnamefont {J.~B.}\ \bibnamefont {Bauer}}, \bibinfo
  {author} {\bibfnamefont {T.}~\bibnamefont {Naumann}}, \bibinfo {author}
  {\bibfnamefont {L.~V.}\ \bibnamefont {M{\"{u}}ller}}, \bibinfo {author}
  {\bibfnamefont {F.}~\bibnamefont {S{\"{a}}ttele}}, \bibinfo {author}
  {\bibfnamefont {H.~F.}\ \bibnamefont {Bettinger}}, \bibinfo {author}
  {\bibfnamefont {R.}~\bibnamefont {Tonner-Zech}}, \ and\ \bibinfo {author}
  {\bibfnamefont {J.~M.}\ \bibnamefont {Gottfried}},\ }\href {\doibase
  10.1021/jacs.4c13296} {\bibfield  {journal} {\bibinfo  {journal} {J. Am.
  Chem. Soc.}\ ,\ \bibinfo {pages} {DOI: 10.1021/jacs.4c13296}} (\bibinfo
  {year} {2025})}\BibitemShut {NoStop}%
\bibitem [{\citenamefont {Tavan}\ and\ \citenamefont
  {Schulten}(1987)}]{Tavan:1987gc}%
  \BibitemOpen
  \bibfield  {author} {\bibinfo {author} {\bibfnamefont {P.}~\bibnamefont
  {Tavan}}\ and\ \bibinfo {author} {\bibfnamefont {K.}~\bibnamefont
  {Schulten}},\ }\href {\doibase 10.1103/PhysRevB.36.4337} {\bibfield
  {journal} {\bibinfo  {journal} {Phys. Rev. B}\ }\textbf {\bibinfo {volume}
  {36}},\ \bibinfo {pages} {4337} (\bibinfo {year} {1987})}\BibitemShut
  {NoStop}%
\bibitem [{\citenamefont {Kopczynski}\ \emph {et~al.}(2007)\citenamefont
  {Kopczynski}, \citenamefont {Ehlers}, \citenamefont {Lenzer},\ and\
  \citenamefont {Oum}}]{Kopczynski2007}%
  \BibitemOpen
  \bibfield  {author} {\bibinfo {author} {\bibfnamefont {M.}~\bibnamefont
  {Kopczynski}}, \bibinfo {author} {\bibfnamefont {F.}~\bibnamefont {Ehlers}},
  \bibinfo {author} {\bibfnamefont {T.}~\bibnamefont {Lenzer}}, \ and\ \bibinfo
  {author} {\bibfnamefont {K.}~\bibnamefont {Oum}},\ }\href {\doibase
  10.1021/jp0672252} {\bibfield  {journal} {\bibinfo  {journal} {J. Phys. Chem.
  A}\ }\textbf {\bibinfo {volume} {111}},\ \bibinfo {pages} {5370} (\bibinfo
  {year} {2007})}\BibitemShut {NoStop}%
\bibitem [{\citenamefont {Ostroumov}\ \emph {et~al.}(2013)\citenamefont
  {Ostroumov}, \citenamefont {Mulvaney}, \citenamefont {Cogdell},\ and\
  \citenamefont {Scholes}}]{Ostroumov2013}%
  \BibitemOpen
  \bibfield  {author} {\bibinfo {author} {\bibfnamefont {E.~E.}\ \bibnamefont
  {Ostroumov}}, \bibinfo {author} {\bibfnamefont {R.~M.}\ \bibnamefont
  {Mulvaney}}, \bibinfo {author} {\bibfnamefont {R.~J.}\ \bibnamefont
  {Cogdell}}, \ and\ \bibinfo {author} {\bibfnamefont {G.~D.}\ \bibnamefont
  {Scholes}},\ }\href {\doibase 10.1126/science.1230106} {\bibfield  {journal}
  {\bibinfo  {journal} {Science (80-. ).}\ }\textbf {\bibinfo {volume} {340}},\
  \bibinfo {pages} {52} (\bibinfo {year} {2013})}\BibitemShut {NoStop}%
\bibitem [{\citenamefont {Wang}\ \emph {et~al.}(2005)\citenamefont {Wang},
  \citenamefont {Nakamura}, \citenamefont {Kanematsu}, \citenamefont {Koyama},
  \citenamefont {Nagae}, \citenamefont {Nishio}, \citenamefont {Hashimoto},\
  and\ \citenamefont {Zhang}}]{Wang2005}%
  \BibitemOpen
  \bibfield  {author} {\bibinfo {author} {\bibfnamefont {P.}~\bibnamefont
  {Wang}}, \bibinfo {author} {\bibfnamefont {R.}~\bibnamefont {Nakamura}},
  \bibinfo {author} {\bibfnamefont {Y.}~\bibnamefont {Kanematsu}}, \bibinfo
  {author} {\bibfnamefont {Y.}~\bibnamefont {Koyama}}, \bibinfo {author}
  {\bibfnamefont {H.}~\bibnamefont {Nagae}}, \bibinfo {author} {\bibfnamefont
  {T.}~\bibnamefont {Nishio}}, \bibinfo {author} {\bibfnamefont
  {H.}~\bibnamefont {Hashimoto}}, \ and\ \bibinfo {author} {\bibfnamefont
  {J.-P.}\ \bibnamefont {Zhang}},\ }\href {\doibase
  10.1016/j.cplett.2005.05.037} {\bibfield  {journal} {\bibinfo  {journal}
  {Chem. Phys. Lett.}\ }\textbf {\bibinfo {volume} {410}},\ \bibinfo {pages}
  {108} (\bibinfo {year} {2005})}\BibitemShut {NoStop}%
\bibitem [{\citenamefont {Andreussi}\ \emph {et~al.}(2015)\citenamefont
  {Andreussi}, \citenamefont {Knecht}, \citenamefont {Marian}, \citenamefont
  {Kongsted},\ and\ \citenamefont {Mennucci}}]{Andreussi2015}%
  \BibitemOpen
  \bibfield  {author} {\bibinfo {author} {\bibfnamefont {O.}~\bibnamefont
  {Andreussi}}, \bibinfo {author} {\bibfnamefont {S.}~\bibnamefont {Knecht}},
  \bibinfo {author} {\bibfnamefont {C.~M.}\ \bibnamefont {Marian}}, \bibinfo
  {author} {\bibfnamefont {J.}~\bibnamefont {Kongsted}}, \ and\ \bibinfo
  {author} {\bibfnamefont {B.}~\bibnamefont {Mennucci}},\ }\href {\doibase
  10.1021/ct5011246} {\bibfield  {journal} {\bibinfo  {journal} {J. Chem.
  Theory Comput.}\ }\textbf {\bibinfo {volume} {11}},\ \bibinfo {pages} {655}
  (\bibinfo {year} {2015})}\BibitemShut {NoStop}%
\bibitem [{\citenamefont {Khokhlov}\ and\ \citenamefont
  {Belov}(2020)}]{Khokhlov:2020kn}%
  \BibitemOpen
  \bibfield  {author} {\bibinfo {author} {\bibfnamefont {D.}~\bibnamefont
  {Khokhlov}}\ and\ \bibinfo {author} {\bibfnamefont {A.}~\bibnamefont
  {Belov}},\ }\href {\doibase 10.1021/acs.jpca.0c01678} {\bibfield  {journal}
  {\bibinfo  {journal} {J. Phys. Chem. A}\ }\textbf {\bibinfo {volume} {124}},\
  \bibinfo {pages} {5790} (\bibinfo {year} {2020})}\BibitemShut {NoStop}%
\bibitem [{\citenamefont {G{\"{o}}tze}\ and\ \citenamefont
  {Thiel}(2013)}]{Gotze2013}%
  \BibitemOpen
  \bibfield  {author} {\bibinfo {author} {\bibfnamefont {J.~P.}\ \bibnamefont
  {G{\"{o}}tze}}\ and\ \bibinfo {author} {\bibfnamefont {W.}~\bibnamefont
  {Thiel}},\ }\href {\doibase 10.1016/j.chemphys.2013.01.030} {\bibfield
  {journal} {\bibinfo  {journal} {Chem. Phys.}\ }\textbf {\bibinfo {volume}
  {415}},\ \bibinfo {pages} {247} (\bibinfo {year} {2013})}\BibitemShut
  {NoStop}%
\bibitem [{\citenamefont {Scuseria}\ and\ \citenamefont
  {Schaefer}(1990)}]{Scuseria1990}%
  \BibitemOpen
  \bibfield  {author} {\bibinfo {author} {\bibfnamefont {G.~E.}\ \bibnamefont
  {Scuseria}}\ and\ \bibinfo {author} {\bibfnamefont {H.~F.}\ \bibnamefont
  {Schaefer}},\ }\href {\doibase 10.1016/S0009-2614(90)87186-U} {\bibfield
  {journal} {\bibinfo  {journal} {Chem. Phys. Lett.}\ }\textbf {\bibinfo
  {volume} {174}},\ \bibinfo {pages} {501} (\bibinfo {year}
  {1990})}\BibitemShut {NoStop}%
\bibitem [{\citenamefont {Roos}\ and\ \citenamefont
  {Andersson}(1995)}]{Roos:1995jz}%
  \BibitemOpen
  \bibfield  {author} {\bibinfo {author} {\bibfnamefont {B.~O.}\ \bibnamefont
  {Roos}}\ and\ \bibinfo {author} {\bibfnamefont {K.}~\bibnamefont
  {Andersson}},\ }\href {\doibase 10.1016/0009-2614(95)01010-7} {\bibfield
  {journal} {\bibinfo  {journal} {Chem. Phys. Lett.}\ }\textbf {\bibinfo
  {volume} {245}},\ \bibinfo {pages} {215} (\bibinfo {year}
  {1995})}\BibitemShut {NoStop}%
\bibitem [{\citenamefont {Angeli}\ \emph {et~al.}(2006)\citenamefont {Angeli},
  \citenamefont {Bories}, \citenamefont {Cavallini},\ and\ \citenamefont
  {Cimiraglia}}]{Angeli:2006gf}%
  \BibitemOpen
  \bibfield  {author} {\bibinfo {author} {\bibfnamefont {C.}~\bibnamefont
  {Angeli}}, \bibinfo {author} {\bibfnamefont {B.}~\bibnamefont {Bories}},
  \bibinfo {author} {\bibfnamefont {A.}~\bibnamefont {Cavallini}}, \ and\
  \bibinfo {author} {\bibfnamefont {R.}~\bibnamefont {Cimiraglia}},\ }\href
  {\doibase 10.1063/1.2148946} {\bibfield  {journal} {\bibinfo  {journal} {J.
  Chem. Phys.}\ }\textbf {\bibinfo {volume} {124}},\ \bibinfo {pages} {054108}
  (\bibinfo {year} {2006})}\BibitemShut {NoStop}%
\bibitem [{\citenamefont {M{\"{u}}ller}(2009)}]{Muller:2009jd}%
  \BibitemOpen
  \bibfield  {author} {\bibinfo {author} {\bibfnamefont {T.}~\bibnamefont
  {M{\"{u}}ller}},\ }\href {\doibase 10.1021/jp905254u} {\bibfield  {journal}
  {\bibinfo  {journal} {J. Phys. Chem. A}\ }\textbf {\bibinfo {volume} {113}},\
  \bibinfo {pages} {12729} (\bibinfo {year} {2009})}\BibitemShut {NoStop}%
\bibitem [{\citenamefont {Li}\ \emph {et~al.}(2020)\citenamefont {Li},
  \citenamefont {Yao}, \citenamefont {Holmes}, \citenamefont {Otten},
  \citenamefont {Sun}, \citenamefont {Sharma},\ and\ \citenamefont
  {Umrigar}}]{Li:2019wp}%
  \BibitemOpen
  \bibfield  {author} {\bibinfo {author} {\bibfnamefont {J.}~\bibnamefont
  {Li}}, \bibinfo {author} {\bibfnamefont {Y.}~\bibnamefont {Yao}}, \bibinfo
  {author} {\bibfnamefont {A.~A.}\ \bibnamefont {Holmes}}, \bibinfo {author}
  {\bibfnamefont {M.}~\bibnamefont {Otten}}, \bibinfo {author} {\bibfnamefont
  {Q.}~\bibnamefont {Sun}}, \bibinfo {author} {\bibfnamefont {S.}~\bibnamefont
  {Sharma}}, \ and\ \bibinfo {author} {\bibfnamefont {C.~J.}\ \bibnamefont
  {Umrigar}},\ }\href {\doibase 10.1103/PhysRevResearch.2.012015} {\bibfield
  {journal} {\bibinfo  {journal} {Phys. Rev. Res.}\ }\textbf {\bibinfo {volume}
  {2}},\ \bibinfo {pages} {012015} (\bibinfo {year} {2020})},\ \Eprint
  {http://arxiv.org/abs/1909.12319} {arXiv:1909.12319} \BibitemShut {NoStop}%
\bibitem [{\citenamefont {Casey}\ and\ \citenamefont
  {Leopold}(1993)}]{Casey:1993gq}%
  \BibitemOpen
  \bibfield  {author} {\bibinfo {author} {\bibfnamefont {S.~M.}\ \bibnamefont
  {Casey}}\ and\ \bibinfo {author} {\bibfnamefont {D.~G.}\ \bibnamefont
  {Leopold}},\ }\href {\doibase 10.1021/j100106a005} {\bibfield  {journal}
  {\bibinfo  {journal} {J. Phys. Chem.}\ }\textbf {\bibinfo {volume} {97}},\
  \bibinfo {pages} {816} (\bibinfo {year} {1993})}\BibitemShut {NoStop}%
\bibitem [{\citenamefont {White}(1993)}]{White:1993fb}%
  \BibitemOpen
  \bibfield  {author} {\bibinfo {author} {\bibfnamefont {S.~R.}\ \bibnamefont
  {White}},\ }\href {\doibase 10.1103/PhysRevB.48.10345} {\bibfield  {journal}
  {\bibinfo  {journal} {Phys. Rev. B}\ }\textbf {\bibinfo {volume} {48}},\
  \bibinfo {pages} {10345} (\bibinfo {year} {1993})}\BibitemShut {NoStop}%
\bibitem [{\citenamefont {Dorando}, \citenamefont {Hachmann},\ and\
  \citenamefont {Chan}(2007)}]{Dorando2007}%
  \BibitemOpen
  \bibfield  {author} {\bibinfo {author} {\bibfnamefont {J.~J.}\ \bibnamefont
  {Dorando}}, \bibinfo {author} {\bibfnamefont {J.}~\bibnamefont {Hachmann}}, \
  and\ \bibinfo {author} {\bibfnamefont {G.~K.-L.}\ \bibnamefont {Chan}},\
  }\href {\doibase 10.1063/1.2768360} {\bibfield  {journal} {\bibinfo
  {journal} {J. Chem. Phys.}\ }\textbf {\bibinfo {volume} {127}},\ \bibinfo
  {pages} {084109} (\bibinfo {year} {2007})},\ \Eprint
  {http://arxiv.org/abs/0707.3121} {arXiv:0707.3121} \BibitemShut {NoStop}%
\bibitem [{\citenamefont {Sharma}\ and\ \citenamefont
  {Chan}(2012)}]{Sharma2012}%
  \BibitemOpen
  \bibfield  {author} {\bibinfo {author} {\bibfnamefont {S.}~\bibnamefont
  {Sharma}}\ and\ \bibinfo {author} {\bibfnamefont {G.~K.-L.}\ \bibnamefont
  {Chan}},\ }\href {\doibase 10.1063/1.3695642} {\bibfield  {journal} {\bibinfo
   {journal} {J. Chem. Phys.}\ }\textbf {\bibinfo {volume} {136}},\ \bibinfo
  {pages} {124121} (\bibinfo {year} {2012})},\ \Eprint
  {http://arxiv.org/abs/1408.5039} {arXiv:1408.5039} \BibitemShut {NoStop}%
\bibitem [{\citenamefont {Wouters}\ \emph {et~al.}(2014)\citenamefont
  {Wouters}, \citenamefont {Poelmans}, \citenamefont {Ayers},\ and\
  \citenamefont {{Van Neck}}}]{Wouters:2014gl}%
  \BibitemOpen
  \bibfield  {author} {\bibinfo {author} {\bibfnamefont {S.}~\bibnamefont
  {Wouters}}, \bibinfo {author} {\bibfnamefont {W.}~\bibnamefont {Poelmans}},
  \bibinfo {author} {\bibfnamefont {P.~W.}\ \bibnamefont {Ayers}}, \ and\
  \bibinfo {author} {\bibfnamefont {D.}~\bibnamefont {{Van Neck}}},\ }\href
  {\doibase 10.1016/j.cpc.2014.01.019} {\bibfield  {journal} {\bibinfo
  {journal} {Comput. Phys. Commun.}\ }\textbf {\bibinfo {volume} {185}},\
  \bibinfo {pages} {1501} (\bibinfo {year} {2014})}\BibitemShut {NoStop}%
\bibitem [{\citenamefont {Keller}\ and\ \citenamefont
  {Reiher}(2016)}]{Keller2016}%
  \BibitemOpen
  \bibfield  {author} {\bibinfo {author} {\bibfnamefont {S.}~\bibnamefont
  {Keller}}\ and\ \bibinfo {author} {\bibfnamefont {M.}~\bibnamefont
  {Reiher}},\ }\href {\doibase 10.1063/1.4944921} {\bibfield  {journal}
  {\bibinfo  {journal} {J. Chem. Phys.}\ }\textbf {\bibinfo {volume} {144}},\
  \bibinfo {pages} {134101} (\bibinfo {year} {2016})},\ \Eprint
  {http://arxiv.org/abs/1602.01145} {arXiv:1602.01145} \BibitemShut {NoStop}%
\bibitem [{\citenamefont {Zhai}\ \emph {et~al.}(2023)\citenamefont {Zhai},
  \citenamefont {Larsson}, \citenamefont {Lee}, \citenamefont {Cui},
  \citenamefont {Zhu}, \citenamefont {Sun}, \citenamefont {Peng}, \citenamefont
  {Peng}, \citenamefont {Liao}, \citenamefont {T{\"{o}}lle}, \citenamefont
  {Yang}, \citenamefont {Li},\ and\ \citenamefont {Chan}}]{Zhai2023}%
  \BibitemOpen
  \bibfield  {author} {\bibinfo {author} {\bibfnamefont {H.}~\bibnamefont
  {Zhai}}, \bibinfo {author} {\bibfnamefont {H.~R.}\ \bibnamefont {Larsson}},
  \bibinfo {author} {\bibfnamefont {S.}~\bibnamefont {Lee}}, \bibinfo {author}
  {\bibfnamefont {Z.-H.}\ \bibnamefont {Cui}}, \bibinfo {author} {\bibfnamefont
  {T.}~\bibnamefont {Zhu}}, \bibinfo {author} {\bibfnamefont {C.}~\bibnamefont
  {Sun}}, \bibinfo {author} {\bibfnamefont {L.}~\bibnamefont {Peng}}, \bibinfo
  {author} {\bibfnamefont {R.}~\bibnamefont {Peng}}, \bibinfo {author}
  {\bibfnamefont {K.}~\bibnamefont {Liao}}, \bibinfo {author} {\bibfnamefont
  {J.}~\bibnamefont {T{\"{o}}lle}}, \bibinfo {author} {\bibfnamefont
  {J.}~\bibnamefont {Yang}}, \bibinfo {author} {\bibfnamefont {S.}~\bibnamefont
  {Li}}, \ and\ \bibinfo {author} {\bibfnamefont {G.~K.-L.}\ \bibnamefont
  {Chan}},\ }\href {\doibase 10.1063/5.0180424} {\bibfield  {journal} {\bibinfo
   {journal} {J. Chem. Phys.}\ }\textbf {\bibinfo {volume} {159}},\ \bibinfo
  {pages} {234801} (\bibinfo {year} {2023})}\BibitemShut {NoStop}%
\bibitem [{\citenamefont {Zgid}\ \emph {et~al.}(2009)\citenamefont {Zgid},
  \citenamefont {Ghosh}, \citenamefont {Neuscamman},\ and\ \citenamefont
  {Chan}}]{Zgid:2009fu}%
  \BibitemOpen
  \bibfield  {author} {\bibinfo {author} {\bibfnamefont {D.}~\bibnamefont
  {Zgid}}, \bibinfo {author} {\bibfnamefont {D.}~\bibnamefont {Ghosh}},
  \bibinfo {author} {\bibfnamefont {E.}~\bibnamefont {Neuscamman}}, \ and\
  \bibinfo {author} {\bibfnamefont {G.~K.-L.}\ \bibnamefont {Chan}},\ }\href
  {\doibase 10.1063/1.3132922} {\bibfield  {journal} {\bibinfo  {journal} {J.
  Chem. Phys.}\ }\textbf {\bibinfo {volume} {130}},\ \bibinfo {pages} {194107}
  (\bibinfo {year} {2009})}\BibitemShut {NoStop}%
\bibitem [{\citenamefont {Li}, \citenamefont {Misiewicz},\ and\ \citenamefont
  {Evangelista}(2023)}]{Li2023a}%
  \BibitemOpen
  \bibfield  {author} {\bibinfo {author} {\bibfnamefont {S.}~\bibnamefont
  {Li}}, \bibinfo {author} {\bibfnamefont {J.~P.}\ \bibnamefont {Misiewicz}}, \
  and\ \bibinfo {author} {\bibfnamefont {F.~A.}\ \bibnamefont {Evangelista}},\
  }\href {\doibase 10.1063/5.0159403} {\bibfield  {journal} {\bibinfo
  {journal} {J. Chem. Phys.}\ }\textbf {\bibinfo {volume} {159}},\ \bibinfo
  {pages} {114106} (\bibinfo {year} {2023})}\BibitemShut {NoStop}%
\bibitem [{\citenamefont {Chatterjee}\ and\ \citenamefont
  {Sokolov}(2020)}]{Chatterjee2020}%
  \BibitemOpen
  \bibfield  {author} {\bibinfo {author} {\bibfnamefont {K.}~\bibnamefont
  {Chatterjee}}\ and\ \bibinfo {author} {\bibfnamefont {A.~Y.}\ \bibnamefont
  {Sokolov}},\ }\href {\doibase 10.1021/acs.jctc.0c00778} {\bibfield  {journal}
  {\bibinfo  {journal} {J. Chem. Theory Comput.}\ }\textbf {\bibinfo {volume}
  {16}},\ \bibinfo {pages} {6343} (\bibinfo {year} {2020})}\BibitemShut
  {NoStop}%
\bibitem [{\citenamefont {Kollmar}\ \emph {et~al.}(2021)\citenamefont
  {Kollmar}, \citenamefont {Sivalingam}, \citenamefont {Guo},\ and\
  \citenamefont {Neese}}]{Kollmar2021}%
  \BibitemOpen
  \bibfield  {author} {\bibinfo {author} {\bibfnamefont {C.}~\bibnamefont
  {Kollmar}}, \bibinfo {author} {\bibfnamefont {K.}~\bibnamefont {Sivalingam}},
  \bibinfo {author} {\bibfnamefont {Y.}~\bibnamefont {Guo}}, \ and\ \bibinfo
  {author} {\bibfnamefont {F.}~\bibnamefont {Neese}},\ }\href {\doibase
  10.1063/5.0072129} {\bibfield  {journal} {\bibinfo  {journal} {J. Chem.
  Phys.}\ }\textbf {\bibinfo {volume} {155}},\ \bibinfo {pages} {234104}
  (\bibinfo {year} {2021})}\BibitemShut {NoStop}%
\bibitem [{\citenamefont {Evangelista}\ \emph {et~al.}(2024)\citenamefont
  {Evangelista}, \citenamefont {Li}, \citenamefont {Verma}, \citenamefont
  {Hannon}, \citenamefont {Schriber}, \citenamefont {Zhang}, \citenamefont
  {Cai}, \citenamefont {Wang}, \citenamefont {He}, \citenamefont {Stair},
  \citenamefont {Huang}, \citenamefont {Huang}, \citenamefont {Misiewicz},
  \citenamefont {Li}, \citenamefont {Marin}, \citenamefont {Zhao},\ and\
  \citenamefont {Burns}}]{Evangelista2024}%
  \BibitemOpen
  \bibfield  {author} {\bibinfo {author} {\bibfnamefont {F.~A.}\ \bibnamefont
  {Evangelista}}, \bibinfo {author} {\bibfnamefont {C.}~\bibnamefont {Li}},
  \bibinfo {author} {\bibfnamefont {P.}~\bibnamefont {Verma}}, \bibinfo
  {author} {\bibfnamefont {K.~P.}\ \bibnamefont {Hannon}}, \bibinfo {author}
  {\bibfnamefont {J.~B.}\ \bibnamefont {Schriber}}, \bibinfo {author}
  {\bibfnamefont {T.}~\bibnamefont {Zhang}}, \bibinfo {author} {\bibfnamefont
  {C.}~\bibnamefont {Cai}}, \bibinfo {author} {\bibfnamefont {S.}~\bibnamefont
  {Wang}}, \bibinfo {author} {\bibfnamefont {N.}~\bibnamefont {He}}, \bibinfo
  {author} {\bibfnamefont {N.~H.}\ \bibnamefont {Stair}}, \bibinfo {author}
  {\bibfnamefont {M.}~\bibnamefont {Huang}}, \bibinfo {author} {\bibfnamefont
  {R.}~\bibnamefont {Huang}}, \bibinfo {author} {\bibfnamefont {J.~P.}\
  \bibnamefont {Misiewicz}}, \bibinfo {author} {\bibfnamefont {S.}~\bibnamefont
  {Li}}, \bibinfo {author} {\bibfnamefont {K.}~\bibnamefont {Marin}}, \bibinfo
  {author} {\bibfnamefont {Z.}~\bibnamefont {Zhao}}, \ and\ \bibinfo {author}
  {\bibfnamefont {L.~A.}\ \bibnamefont {Burns}},\ }\href {\doibase
  10.1063/5.0216512} {\bibfield  {journal} {\bibinfo  {journal} {J. Chem.
  Phys.}\ }\textbf {\bibinfo {volume} {161}},\ \bibinfo {pages} {062502}
  (\bibinfo {year} {2024})}\BibitemShut {NoStop}%
\bibitem [{MyF(2025)}]{MyForte}%
  \BibitemOpen
  \href@noop {} {}\bibinfo {howpublished} {See
  \url{https://github.com/lcyyork/forte/tree/dmrg_init_orbs}, last accessed:
  Feb.~17} (\bibinfo {year} {2025})\BibitemShut {NoStop}%
\bibitem [{\citenamefont {Sharma}\ \emph {et~al.}(2019)\citenamefont {Sharma},
  \citenamefont {Bernales}, \citenamefont {Knecht}, \citenamefont {Truhlar},\
  and\ \citenamefont {Gagliardi}}]{Sharma2019}%
  \BibitemOpen
  \bibfield  {author} {\bibinfo {author} {\bibfnamefont {P.}~\bibnamefont
  {Sharma}}, \bibinfo {author} {\bibfnamefont {V.}~\bibnamefont {Bernales}},
  \bibinfo {author} {\bibfnamefont {S.}~\bibnamefont {Knecht}}, \bibinfo
  {author} {\bibfnamefont {D.~G.}\ \bibnamefont {Truhlar}}, \ and\ \bibinfo
  {author} {\bibfnamefont {L.}~\bibnamefont {Gagliardi}},\ }\href {\doibase
  10.1039/C8SC03569E} {\bibfield  {journal} {\bibinfo  {journal} {Chem. Sci.}\
  }\textbf {\bibinfo {volume} {10}},\ \bibinfo {pages} {1716} (\bibinfo {year}
  {2019})},\ \Eprint {http://arxiv.org/abs/1808.06273} {arXiv:1808.06273}
  \BibitemShut {NoStop}%
\bibitem [{\citenamefont {Pipek}\ and\ \citenamefont
  {Mezey}(1989)}]{Pipek:1989ci}%
  \BibitemOpen
  \bibfield  {author} {\bibinfo {author} {\bibfnamefont {J.}~\bibnamefont
  {Pipek}}\ and\ \bibinfo {author} {\bibfnamefont {P.~G.}\ \bibnamefont
  {Mezey}},\ }\href {\doibase 10.1063/1.456588} {\bibfield  {journal} {\bibinfo
   {journal} {J. Chem. Phys.}\ }\textbf {\bibinfo {volume} {90}},\ \bibinfo
  {pages} {4916} (\bibinfo {year} {1989})}\BibitemShut {NoStop}%
\bibitem [{\citenamefont {Olivares-Amaya}\ \emph {et~al.}(2015)\citenamefont
  {Olivares-Amaya}, \citenamefont {Hu}, \citenamefont {Nakatani}, \citenamefont
  {Sharma}, \citenamefont {Yang},\ and\ \citenamefont
  {Chan}}]{OlivaresAmaya:2015hi}%
  \BibitemOpen
  \bibfield  {author} {\bibinfo {author} {\bibfnamefont {R.}~\bibnamefont
  {Olivares-Amaya}}, \bibinfo {author} {\bibfnamefont {W.}~\bibnamefont {Hu}},
  \bibinfo {author} {\bibfnamefont {N.}~\bibnamefont {Nakatani}}, \bibinfo
  {author} {\bibfnamefont {S.}~\bibnamefont {Sharma}}, \bibinfo {author}
  {\bibfnamefont {J.}~\bibnamefont {Yang}}, \ and\ \bibinfo {author}
  {\bibfnamefont {G.~K.-L.}\ \bibnamefont {Chan}},\ }\href {\doibase
  10.1063/1.4905329} {\bibfield  {journal} {\bibinfo  {journal} {J. Chem.
  Phys.}\ }\textbf {\bibinfo {volume} {142}},\ \bibinfo {pages} {034102}
  (\bibinfo {year} {2015})}\BibitemShut {NoStop}%
\bibitem [{\citenamefont {Kurashige}\ and\ \citenamefont
  {Yanai}(2009)}]{Kurashige:2009gs}%
  \BibitemOpen
  \bibfield  {author} {\bibinfo {author} {\bibfnamefont {Y.}~\bibnamefont
  {Kurashige}}\ and\ \bibinfo {author} {\bibfnamefont {T.}~\bibnamefont
  {Yanai}},\ }\href {\doibase 10.1063/1.3152576} {\bibfield  {journal}
  {\bibinfo  {journal} {J. Chem. Phys.}\ }\textbf {\bibinfo {volume} {130}},\
  \bibinfo {pages} {234114} (\bibinfo {year} {2009})}\BibitemShut {NoStop}%
\bibitem [{\citenamefont {Hess}(1986)}]{Hess:1986cz}%
  \BibitemOpen
  \bibfield  {author} {\bibinfo {author} {\bibfnamefont {B.~A.}\ \bibnamefont
  {Hess}},\ }\href@noop {} {\bibfield  {journal} {\bibinfo  {journal} {Phys.
  Rev. A}\ }\textbf {\bibinfo {volume} {33}},\ \bibinfo {pages} {3742}
  (\bibinfo {year} {1986})}\BibitemShut {NoStop}%
\bibitem [{\citenamefont {Wolf}, \citenamefont {Reiher},\ and\ \citenamefont
  {Hess}(2002)}]{Wolf:2002ia}%
  \BibitemOpen
  \bibfield  {author} {\bibinfo {author} {\bibfnamefont {A.}~\bibnamefont
  {Wolf}}, \bibinfo {author} {\bibfnamefont {M.}~\bibnamefont {Reiher}}, \ and\
  \bibinfo {author} {\bibfnamefont {B.~A.}\ \bibnamefont {Hess}},\ }\href
  {\doibase 10.1063/1.1515314} {\bibfield  {journal} {\bibinfo  {journal} {J.
  Chem. Phys.}\ }\textbf {\bibinfo {volume} {117}},\ \bibinfo {pages} {9215}
  (\bibinfo {year} {2002})}\BibitemShut {NoStop}%
\bibitem [{\citenamefont {Feller}(1993)}]{Feller:1993ex}%
  \BibitemOpen
  \bibfield  {author} {\bibinfo {author} {\bibfnamefont {D.}~\bibnamefont
  {Feller}},\ }\href@noop {} {\bibfield  {journal} {\bibinfo  {journal} {J.
  Chem. Phys.}\ }\textbf {\bibinfo {volume} {98}},\ \bibinfo {pages} {7059}
  (\bibinfo {year} {1993})}\BibitemShut {NoStop}%
\bibitem [{\citenamefont {Helgaker}\ \emph {et~al.}(1997)\citenamefont
  {Helgaker}, \citenamefont {Klopper}, \citenamefont {Koch},\ and\
  \citenamefont {Noga}}]{Helgaker:1997bb}%
  \BibitemOpen
  \bibfield  {author} {\bibinfo {author} {\bibfnamefont {T.}~\bibnamefont
  {Helgaker}}, \bibinfo {author} {\bibfnamefont {W.}~\bibnamefont {Klopper}},
  \bibinfo {author} {\bibfnamefont {H.}~\bibnamefont {Koch}}, \ and\ \bibinfo
  {author} {\bibfnamefont {J.}~\bibnamefont {Noga}},\ }\href@noop {} {\bibfield
   {journal} {\bibinfo  {journal} {J. Chem. Phys.}\ }\textbf {\bibinfo {volume}
  {106}},\ \bibinfo {pages} {9639} (\bibinfo {year} {1997})}\BibitemShut
  {NoStop}%
\bibitem [{\citenamefont {Balabanov}\ and\ \citenamefont
  {Peterson}(2005)}]{Balabanov:2005hm}%
  \BibitemOpen
  \bibfield  {author} {\bibinfo {author} {\bibfnamefont {N.~B.}\ \bibnamefont
  {Balabanov}}\ and\ \bibinfo {author} {\bibfnamefont {K.~A.}\ \bibnamefont
  {Peterson}},\ }\href {\doibase 10.1063/1.1998907} {\bibfield  {journal}
  {\bibinfo  {journal} {J. Chem. Phys.}\ }\textbf {\bibinfo {volume} {123}},\
  \bibinfo {pages} {064107} (\bibinfo {year} {2005})}\BibitemShut {NoStop}%
\bibitem [{\citenamefont {Smith}\ \emph {et~al.}(2020)\citenamefont {Smith},
  \citenamefont {Burns}, \citenamefont {Simmonett}, \citenamefont {Parrish},
  \citenamefont {Schieber}, \citenamefont {Galvelis}, \citenamefont {Kraus},
  \citenamefont {Kruse}, \citenamefont {{Di Remigio}}, \citenamefont
  {Alenaizan}, \citenamefont {James}, \citenamefont {Lehtola}, \citenamefont
  {Misiewicz}, \citenamefont {Scheurer}, \citenamefont {Shaw}, \citenamefont
  {Schriber}, \citenamefont {Xie}, \citenamefont {Glick}, \citenamefont
  {Sirianni}, \citenamefont {O'Brien}, \citenamefont {Waldrop}, \citenamefont
  {Kumar}, \citenamefont {Hohenstein}, \citenamefont {Pritchard}, \citenamefont
  {Brooks}, \citenamefont {Schaefer}, \citenamefont {Sokolov}, \citenamefont
  {Patkowski}, \citenamefont {DePrince}, \citenamefont {Bozkaya}, \citenamefont
  {King}, \citenamefont {Evangelista}, \citenamefont {Turney}, \citenamefont
  {Crawford},\ and\ \citenamefont {Sherrill}}]{Smith:2020ci}%
  \BibitemOpen
  \bibfield  {author} {\bibinfo {author} {\bibfnamefont {D.~G.~A.}\
  \bibnamefont {Smith}}, \bibinfo {author} {\bibfnamefont {L.~A.}\ \bibnamefont
  {Burns}}, \bibinfo {author} {\bibfnamefont {A.~C.}\ \bibnamefont
  {Simmonett}}, \bibinfo {author} {\bibfnamefont {R.~M.}\ \bibnamefont
  {Parrish}}, \bibinfo {author} {\bibfnamefont {M.~C.}\ \bibnamefont
  {Schieber}}, \bibinfo {author} {\bibfnamefont {R.}~\bibnamefont {Galvelis}},
  \bibinfo {author} {\bibfnamefont {P.}~\bibnamefont {Kraus}}, \bibinfo
  {author} {\bibfnamefont {H.}~\bibnamefont {Kruse}}, \bibinfo {author}
  {\bibfnamefont {R.}~\bibnamefont {{Di Remigio}}}, \bibinfo {author}
  {\bibfnamefont {A.}~\bibnamefont {Alenaizan}}, \bibinfo {author}
  {\bibfnamefont {A.~M.}\ \bibnamefont {James}}, \bibinfo {author}
  {\bibfnamefont {S.}~\bibnamefont {Lehtola}}, \bibinfo {author} {\bibfnamefont
  {J.~P.}\ \bibnamefont {Misiewicz}}, \bibinfo {author} {\bibfnamefont
  {M.}~\bibnamefont {Scheurer}}, \bibinfo {author} {\bibfnamefont {R.~A.}\
  \bibnamefont {Shaw}}, \bibinfo {author} {\bibfnamefont {J.~B.}\ \bibnamefont
  {Schriber}}, \bibinfo {author} {\bibfnamefont {Y.}~\bibnamefont {Xie}},
  \bibinfo {author} {\bibfnamefont {Z.~L.}\ \bibnamefont {Glick}}, \bibinfo
  {author} {\bibfnamefont {D.~A.}\ \bibnamefont {Sirianni}}, \bibinfo {author}
  {\bibfnamefont {J.~S.}\ \bibnamefont {O'Brien}}, \bibinfo {author}
  {\bibfnamefont {J.~M.}\ \bibnamefont {Waldrop}}, \bibinfo {author}
  {\bibfnamefont {A.}~\bibnamefont {Kumar}}, \bibinfo {author} {\bibfnamefont
  {E.~G.}\ \bibnamefont {Hohenstein}}, \bibinfo {author} {\bibfnamefont
  {B.~P.}\ \bibnamefont {Pritchard}}, \bibinfo {author} {\bibfnamefont {B.~R.}\
  \bibnamefont {Brooks}}, \bibinfo {author} {\bibfnamefont {H.~F.}\
  \bibnamefont {Schaefer}}, \bibinfo {author} {\bibfnamefont {A.~Y.}\
  \bibnamefont {Sokolov}}, \bibinfo {author} {\bibfnamefont {K.}~\bibnamefont
  {Patkowski}}, \bibinfo {author} {\bibfnamefont {A.~E.}\ \bibnamefont
  {DePrince}}, \bibinfo {author} {\bibfnamefont {U.}~\bibnamefont {Bozkaya}},
  \bibinfo {author} {\bibfnamefont {R.~A.}\ \bibnamefont {King}}, \bibinfo
  {author} {\bibfnamefont {F.~A.}\ \bibnamefont {Evangelista}}, \bibinfo
  {author} {\bibfnamefont {J.~M.}\ \bibnamefont {Turney}}, \bibinfo {author}
  {\bibfnamefont {T.~D.}\ \bibnamefont {Crawford}}, \ and\ \bibinfo {author}
  {\bibfnamefont {C.~D.}\ \bibnamefont {Sherrill}},\ }\href {\doibase
  10.1063/5.0006002} {\bibfield  {journal} {\bibinfo  {journal} {J. Chem.
  Phys.}\ }\textbf {\bibinfo {volume} {152}},\ \bibinfo {pages} {184108}
  (\bibinfo {year} {2020})}\BibitemShut {NoStop}%
\bibitem [{\citenamefont {Pritchard}\ \emph {et~al.}(2019)\citenamefont
  {Pritchard}, \citenamefont {Altarawy}, \citenamefont {Didier}, \citenamefont
  {Gibson},\ and\ \citenamefont {Windus}}]{Pritchard:2019gs}%
  \BibitemOpen
  \bibfield  {author} {\bibinfo {author} {\bibfnamefont {B.~P.}\ \bibnamefont
  {Pritchard}}, \bibinfo {author} {\bibfnamefont {D.}~\bibnamefont {Altarawy}},
  \bibinfo {author} {\bibfnamefont {B.}~\bibnamefont {Didier}}, \bibinfo
  {author} {\bibfnamefont {T.~D.}\ \bibnamefont {Gibson}}, \ and\ \bibinfo
  {author} {\bibfnamefont {T.~L.}\ \bibnamefont {Windus}},\ }\href {\doibase
  10.1021/acs.jcim.9b00725} {\bibfield  {journal} {\bibinfo  {journal} {J.
  Chem. Inf. Model.}\ }\textbf {\bibinfo {volume} {59}},\ \bibinfo {pages}
  {4814} (\bibinfo {year} {2019})}\BibitemShut {NoStop}%
\bibitem [{\citenamefont {Stoychev}, \citenamefont {Auer},\ and\ \citenamefont
  {Neese}(2017)}]{Stoychev2017}%
  \BibitemOpen
  \bibfield  {author} {\bibinfo {author} {\bibfnamefont {G.~L.}\ \bibnamefont
  {Stoychev}}, \bibinfo {author} {\bibfnamefont {A.~A.}\ \bibnamefont {Auer}},
  \ and\ \bibinfo {author} {\bibfnamefont {F.}~\bibnamefont {Neese}},\ }\href
  {\doibase 10.1021/acs.jctc.6b01041} {\bibfield  {journal} {\bibinfo
  {journal} {J. Chem. Theory Comput.}\ }\textbf {\bibinfo {volume} {13}},\
  \bibinfo {pages} {554} (\bibinfo {year} {2017})}\BibitemShut {NoStop}%
\bibitem [{\citenamefont {Li}\ \emph {et~al.}(2024)\citenamefont {Li},
  \citenamefont {Mao}, \citenamefont {Huang},\ and\ \citenamefont
  {Evangelista}}]{Li2024c}%
  \BibitemOpen
  \bibfield  {author} {\bibinfo {author} {\bibfnamefont {C.}~\bibnamefont
  {Li}}, \bibinfo {author} {\bibfnamefont {S.}~\bibnamefont {Mao}}, \bibinfo
  {author} {\bibfnamefont {R.}~\bibnamefont {Huang}}, \ and\ \bibinfo {author}
  {\bibfnamefont {F.~A.}\ \bibnamefont {Evangelista}},\ }\href {\doibase
  10.1021/acs.jctc.4c00152} {\bibfield  {journal} {\bibinfo  {journal} {J.
  Chem. Theory Comput.}\ }\textbf {\bibinfo {volume} {20}},\ \bibinfo {pages}
  {4170} (\bibinfo {year} {2024})}\BibitemShut {NoStop}%
\bibitem [{\citenamefont {Beran}\ \emph {et~al.}(2021)\citenamefont {Beran},
  \citenamefont {Matou{\v{s}}ek}, \citenamefont {Hapka}, \citenamefont
  {Pernal},\ and\ \citenamefont {Veis}}]{Beran2021}%
  \BibitemOpen
  \bibfield  {author} {\bibinfo {author} {\bibfnamefont {P.}~\bibnamefont
  {Beran}}, \bibinfo {author} {\bibfnamefont {M.}~\bibnamefont
  {Matou{\v{s}}ek}}, \bibinfo {author} {\bibfnamefont {M.}~\bibnamefont
  {Hapka}}, \bibinfo {author} {\bibfnamefont {K.}~\bibnamefont {Pernal}}, \
  and\ \bibinfo {author} {\bibfnamefont {L.}~\bibnamefont {Veis}},\ }\href
  {\doibase 10.1021/acs.jctc.1c00896} {\bibfield  {journal} {\bibinfo
  {journal} {J. Chem. Theory Comput.}\ }\textbf {\bibinfo {volume} {17}},\
  \bibinfo {pages} {7575} (\bibinfo {year} {2021})},\ \Eprint
  {http://arxiv.org/abs/2108.12803} {arXiv:2108.12803} \BibitemShut {NoStop}%
\bibitem [{\citenamefont {Hajgat{\'{o}}}, \citenamefont {Huzak},\ and\
  \citenamefont {Deleuze}(2011)}]{Hajgato2011}%
  \BibitemOpen
  \bibfield  {author} {\bibinfo {author} {\bibfnamefont {B.}~\bibnamefont
  {Hajgat{\'{o}}}}, \bibinfo {author} {\bibfnamefont {M.}~\bibnamefont
  {Huzak}}, \ and\ \bibinfo {author} {\bibfnamefont {M.~S.}\ \bibnamefont
  {Deleuze}},\ }\href {\doibase 10.1021/jp2043043} {\bibfield  {journal}
  {\bibinfo  {journal} {J. Phys. Chem. A}\ }\textbf {\bibinfo {volume} {115}},\
  \bibinfo {pages} {9282} (\bibinfo {year} {2011})}\BibitemShut {NoStop}%
\bibitem [{\citenamefont {East}\ and\ \citenamefont
  {Allen}(1993)}]{East:1993cl}%
  \BibitemOpen
  \bibfield  {author} {\bibinfo {author} {\bibfnamefont {A.~L.~L.}\
  \bibnamefont {East}}\ and\ \bibinfo {author} {\bibfnamefont {W.~D.}\
  \bibnamefont {Allen}},\ }\href@noop {} {\bibfield  {journal} {\bibinfo
  {journal} {J. Chem. Phys.}\ }\textbf {\bibinfo {volume} {99}},\ \bibinfo
  {pages} {4638} (\bibinfo {year} {1993})}\BibitemShut {NoStop}%
\bibitem [{\citenamefont {V{\'{e}}ril}\ \emph {et~al.}(2021)\citenamefont
  {V{\'{e}}ril}, \citenamefont {Scemama}, \citenamefont {Caffarel},
  \citenamefont {Lipparini}, \citenamefont {Boggio‐Pasqua}, \citenamefont
  {Jacquemin},\ and\ \citenamefont {Loos}}]{Veril2021}%
  \BibitemOpen
  \bibfield  {author} {\bibinfo {author} {\bibfnamefont {M.}~\bibnamefont
  {V{\'{e}}ril}}, \bibinfo {author} {\bibfnamefont {A.}~\bibnamefont
  {Scemama}}, \bibinfo {author} {\bibfnamefont {M.}~\bibnamefont {Caffarel}},
  \bibinfo {author} {\bibfnamefont {F.}~\bibnamefont {Lipparini}}, \bibinfo
  {author} {\bibfnamefont {M.}~\bibnamefont {Boggio‐Pasqua}}, \bibinfo
  {author} {\bibfnamefont {D.}~\bibnamefont {Jacquemin}}, \ and\ \bibinfo
  {author} {\bibfnamefont {P.}~\bibnamefont {Loos}},\ }\href {\doibase
  10.1002/wcms.1517} {\bibfield  {journal} {\bibinfo  {journal} {WIREs Comput.
  Mol. Sci.}\ }\textbf {\bibinfo {volume} {11}},\ \bibinfo {pages} {e1517}
  (\bibinfo {year} {2021})},\ \Eprint {http://arxiv.org/abs/2011.14675}
  {arXiv:2011.14675} \BibitemShut {NoStop}%
\bibitem [{\citenamefont {Mullinax}\ \emph {et~al.}(2019)\citenamefont
  {Mullinax}, \citenamefont {Epifanovsky}, \citenamefont {Gidofalvi},\ and\
  \citenamefont {DePrince}}]{Mullinax2019}%
  \BibitemOpen
  \bibfield  {author} {\bibinfo {author} {\bibfnamefont {J.~W.}\ \bibnamefont
  {Mullinax}}, \bibinfo {author} {\bibfnamefont {E.}~\bibnamefont
  {Epifanovsky}}, \bibinfo {author} {\bibfnamefont {G.}~\bibnamefont
  {Gidofalvi}}, \ and\ \bibinfo {author} {\bibfnamefont {A.~E.}\ \bibnamefont
  {DePrince}},\ }\href {\doibase 10.1021/acs.jctc.8b00973} {\bibfield
  {journal} {\bibinfo  {journal} {J. Chem. Theory Comput.}\ }\textbf {\bibinfo
  {volume} {15}},\ \bibinfo {pages} {276} (\bibinfo {year} {2019})}\BibitemShut
  {NoStop}%
\bibitem [{\citenamefont {Josue}\ and\ \citenamefont
  {Frank}(2002)}]{Josue2002}%
  \BibitemOpen
  \bibfield  {author} {\bibinfo {author} {\bibfnamefont {J.~S.}\ \bibnamefont
  {Josue}}\ and\ \bibinfo {author} {\bibfnamefont {H.~A.}\ \bibnamefont
  {Frank}},\ }\href {\doibase 10.1021/jp014150n} {\bibfield  {journal}
  {\bibinfo  {journal} {J. Phys. Chem. A}\ }\textbf {\bibinfo {volume} {106}},\
  \bibinfo {pages} {4815} (\bibinfo {year} {2002})}\BibitemShut {NoStop}%
\bibitem [{\citenamefont {Simard}\ \emph {et~al.}(1998)\citenamefont {Simard},
  \citenamefont {Lebeault-Dorget}, \citenamefont {Marijnissen},\ and\
  \citenamefont {ter Meulen}}]{Simard1998}%
  \BibitemOpen
  \bibfield  {author} {\bibinfo {author} {\bibfnamefont {B.}~\bibnamefont
  {Simard}}, \bibinfo {author} {\bibfnamefont {M.-A.}\ \bibnamefont
  {Lebeault-Dorget}}, \bibinfo {author} {\bibfnamefont {A.}~\bibnamefont
  {Marijnissen}}, \ and\ \bibinfo {author} {\bibfnamefont {J.~J.}\ \bibnamefont
  {ter Meulen}},\ }\href {\doibase 10.1063/1.476442} {\bibfield  {journal}
  {\bibinfo  {journal} {J. Chem. Phys.}\ }\textbf {\bibinfo {volume} {108}},\
  \bibinfo {pages} {9668} (\bibinfo {year} {1998})}\BibitemShut {NoStop}%
\bibitem [{\citenamefont {Vancoillie}, \citenamefont {Malmqvist},\ and\
  \citenamefont {Veryazov}(2016)}]{Vancoillie:2016gp}%
  \BibitemOpen
  \bibfield  {author} {\bibinfo {author} {\bibfnamefont {S.}~\bibnamefont
  {Vancoillie}}, \bibinfo {author} {\bibfnamefont {P.~{\AA}.}\ \bibnamefont
  {Malmqvist}}, \ and\ \bibinfo {author} {\bibfnamefont {V.}~\bibnamefont
  {Veryazov}},\ }\href {\doibase 10.1021/acs.jctc.6b00034} {\bibfield
  {journal} {\bibinfo  {journal} {J. Chem. Theory Comput.}\ }\textbf {\bibinfo
  {volume} {12}},\ \bibinfo {pages} {1647} (\bibinfo {year}
  {2016})}\BibitemShut {NoStop}%
\bibitem [{\citenamefont {Lei}\ \emph {et~al.}(2025)\citenamefont {Lei},
  \citenamefont {Guo}, \citenamefont {Suo},\ and\ \citenamefont
  {Liu}}]{Lei2025}%
  \BibitemOpen
  \bibfield  {author} {\bibinfo {author} {\bibfnamefont {Y.}~\bibnamefont
  {Lei}}, \bibinfo {author} {\bibfnamefont {Y.}~\bibnamefont {Guo}}, \bibinfo
  {author} {\bibfnamefont {B.}~\bibnamefont {Suo}}, \ and\ \bibinfo {author}
  {\bibfnamefont {W.}~\bibnamefont {Liu}},\ }\href {\doibase
  10.1021/acs.jctc.4c01596} {\bibfield  {journal} {\bibinfo  {journal} {J.
  Chem. Theory Comput.}\ ,\ \bibinfo {pages} {DOI: 10.1021/acs.jctc.4c01596}}
  (\bibinfo {year} {2025})}\BibitemShut {NoStop}%
\bibitem [{\citenamefont {Bender}\ \emph {et~al.}(2014)\citenamefont {Bender},
  \citenamefont {Doraiswamy}, \citenamefont {Truhlar},\ and\ \citenamefont
  {Candler}}]{Bender:2014do}%
  \BibitemOpen
  \bibfield  {author} {\bibinfo {author} {\bibfnamefont {J.~D.}\ \bibnamefont
  {Bender}}, \bibinfo {author} {\bibfnamefont {S.}~\bibnamefont {Doraiswamy}},
  \bibinfo {author} {\bibfnamefont {D.~G.}\ \bibnamefont {Truhlar}}, \ and\
  \bibinfo {author} {\bibfnamefont {G.~V.}\ \bibnamefont {Candler}},\ }\href
  {\doibase 10.1063/1.4862157} {\bibfield  {journal} {\bibinfo  {journal} {J.
  Chem. Phys.}\ }\textbf {\bibinfo {volume} {140}},\ \bibinfo {pages} {54302}
  (\bibinfo {year} {2014})}\BibitemShut {NoStop}%
\bibitem [{\citenamefont {Hilpert}\ and\ \citenamefont
  {Ruthardt}(1987)}]{Hilpert1987}%
  \BibitemOpen
  \bibfield  {author} {\bibinfo {author} {\bibfnamefont {K.}~\bibnamefont
  {Hilpert}}\ and\ \bibinfo {author} {\bibfnamefont {R.}~\bibnamefont
  {Ruthardt}},\ }\href {\doibase 10.1002/bbpc.19870910707} {\bibfield
  {journal} {\bibinfo  {journal} {Berichte der Bunsengesellschaft f{\"{u}}r
  Phys. Chemie}\ }\textbf {\bibinfo {volume} {91}},\ \bibinfo {pages} {724}
  (\bibinfo {year} {1987})}\BibitemShut {NoStop}%
\bibitem [{\citenamefont {Bondybey}\ and\ \citenamefont
  {English}(1983)}]{Bondybey1983}%
  \BibitemOpen
  \bibfield  {author} {\bibinfo {author} {\bibfnamefont {V.}~\bibnamefont
  {Bondybey}}\ and\ \bibinfo {author} {\bibfnamefont {J.}~\bibnamefont
  {English}},\ }\href {\doibase 10.1016/0009-2614(83)85029-5} {\bibfield
  {journal} {\bibinfo  {journal} {Chem. Phys. Lett.}\ }\textbf {\bibinfo
  {volume} {94}},\ \bibinfo {pages} {443} (\bibinfo {year} {1983})}\BibitemShut
  {NoStop}%
\bibitem [{\citenamefont {Wang}, \citenamefont {Li},\ and\ \citenamefont
  {Evangelista}(2021)}]{Wang2021}%
  \BibitemOpen
  \bibfield  {author} {\bibinfo {author} {\bibfnamefont {S.}~\bibnamefont
  {Wang}}, \bibinfo {author} {\bibfnamefont {C.}~\bibnamefont {Li}}, \ and\
  \bibinfo {author} {\bibfnamefont {F.~A.}\ \bibnamefont {Evangelista}},\
  }\href {\doibase 10.1021/acs.jctc.1c00980} {\bibfield  {journal} {\bibinfo
  {journal} {J. Chem. Theory Comput.}\ }\textbf {\bibinfo {volume} {17}},\
  \bibinfo {pages} {7666} (\bibinfo {year} {2021})},\ \Eprint
  {http://arxiv.org/abs/2109.14548} {arXiv:2109.14548} \BibitemShut {NoStop}%
\bibitem [{\citenamefont {Park}(2022)}]{Park2021c}%
  \BibitemOpen
  \bibfield  {author} {\bibinfo {author} {\bibfnamefont {J.~W.}\ \bibnamefont
  {Park}},\ }\href {\doibase 10.1021/acs.jctc.1c01150} {\bibfield  {journal}
  {\bibinfo  {journal} {J. Chem. Theory Comput.}\ }\textbf {\bibinfo {volume}
  {18}},\ \bibinfo {pages} {2233} (\bibinfo {year} {2022})},\ \Eprint
  {http://arxiv.org/abs/2111.07253} {arXiv:2111.07253} \BibitemShut {NoStop}%
\bibitem [{\citenamefont {Yu}\ and\ \citenamefont {Heine}(2023)}]{Yu2023a}%
  \BibitemOpen
  \bibfield  {author} {\bibinfo {author} {\bibfnamefont {H.}~\bibnamefont
  {Yu}}\ and\ \bibinfo {author} {\bibfnamefont {T.}~\bibnamefont {Heine}},\
  }\href {\doibase 10.1021/jacs.3c05178} {\bibfield  {journal} {\bibinfo
  {journal} {J. Am. Chem. Soc.}\ }\textbf {\bibinfo {volume} {145}},\ \bibinfo
  {pages} {19303} (\bibinfo {year} {2023})}\BibitemShut {NoStop}%
\bibitem [{\citenamefont {Accomasso}\ \emph {et~al.}(2024)\citenamefont
  {Accomasso}, \citenamefont {Londi}, \citenamefont {Cupellini},\ and\
  \citenamefont {Mennucci}}]{Accomasso2024}%
  \BibitemOpen
  \bibfield  {author} {\bibinfo {author} {\bibfnamefont {D.}~\bibnamefont
  {Accomasso}}, \bibinfo {author} {\bibfnamefont {G.}~\bibnamefont {Londi}},
  \bibinfo {author} {\bibfnamefont {L.}~\bibnamefont {Cupellini}}, \ and\
  \bibinfo {author} {\bibfnamefont {B.}~\bibnamefont {Mennucci}},\ }\href
  {\doibase 10.1038/s41467-024-45090-9} {\bibfield  {journal} {\bibinfo
  {journal} {Nat. Commun.}\ }\textbf {\bibinfo {volume} {15}},\ \bibinfo
  {pages} {847} (\bibinfo {year} {2024})}\BibitemShut {NoStop}%
\end{thebibliography}
%

\end{document}